\documentclass{ar-1col}
\usepackage[numbers]{natbib}
\usepackage{url}
\usepackage{subfig}
\usepackage{graphicx}
\graphicspath{ {figures/} }
\newcommand*{\ZOnOn}{\ensuremath{0n0n}}
\newcommand*{\ZXnOn}{\ensuremath{\mathrm{X}n0n}}
\newcommand*{\ZXnXn}{\ensuremath{\mathrm{X}n\mathrm{X}n}}   
\newcommand*{\gamgam}{\mbox{\ensuremath{\gamma\gamma}}}
\newcommand*{\gamongam}{\mbox{\ensuremath{\gamma+\gamma}}}

\newcommand*{\Jpsi}{\ensuremath{J/\psi}}
\newcommand*{\pT}{\ensuremath{p_{\mathrm{T}}}}

\newcommand*{\Pnohad}{\ensuremath{P_\mathrm{0had}}}

\newcommand*{\Pzdc}{\ensuremath{P_\mathrm{fn}}}
\newcommand*{\PY}{\textsc{Pythia}}
\newcommand*{\SL}{\textsc{STARLight}}

\jname{Xxxx. Xxx. Xxx. Xxx.}
\jvol{AA}
\jyear{YYYY}
\doi{10.1146/((please add article doi))}

\begin{document}

% Page header
\markboth{Klein et al.}{Two-photon and Photonuclear reactions}

% Title
\title{Photonuclear and Two-photon Interactions at High-Energy Nuclear Colliders }

%Authors, affiliations address.
\author{Spencer R. Klein$^1$ and Peter Steinberg$^2$
\affil{$^1$Nuclear Science Division, Lawrence Berkeley National Laboratory, Berkeley, CA USA 94720; email: srklein@lbl.gov}
\affil{$^2$Brookhaven National Laboratory, Upton, NY 11973-5000; email: peter.steinberg@bnl.gov}
}

\begin{abstract}
Ultra-peripheral collisions of heavy ions and protons are the energy frontier for electromagnetic interactions.  Both photonuclear and two-photon collisions are studied, at collision energies that are far higher than are available elsewhere.  In this review, we will discuss  physics topics that can be addressed with UPCs, including nuclear shadowing and nuclear structure and searches for beyond-standard-model physics. 
\end{abstract}

\begin{keywords}
ultra-peripheral collisions, photonuclear interactions, two-photon interactions, heavy-ion collisions, vector mesons, nuclear imaging
\end{keywords}
\maketitle

\tableofcontents

\section{Introduction: ultraperipheral collisions}

\begin{figure}[t]
\includegraphics[width=0.9\textwidth]{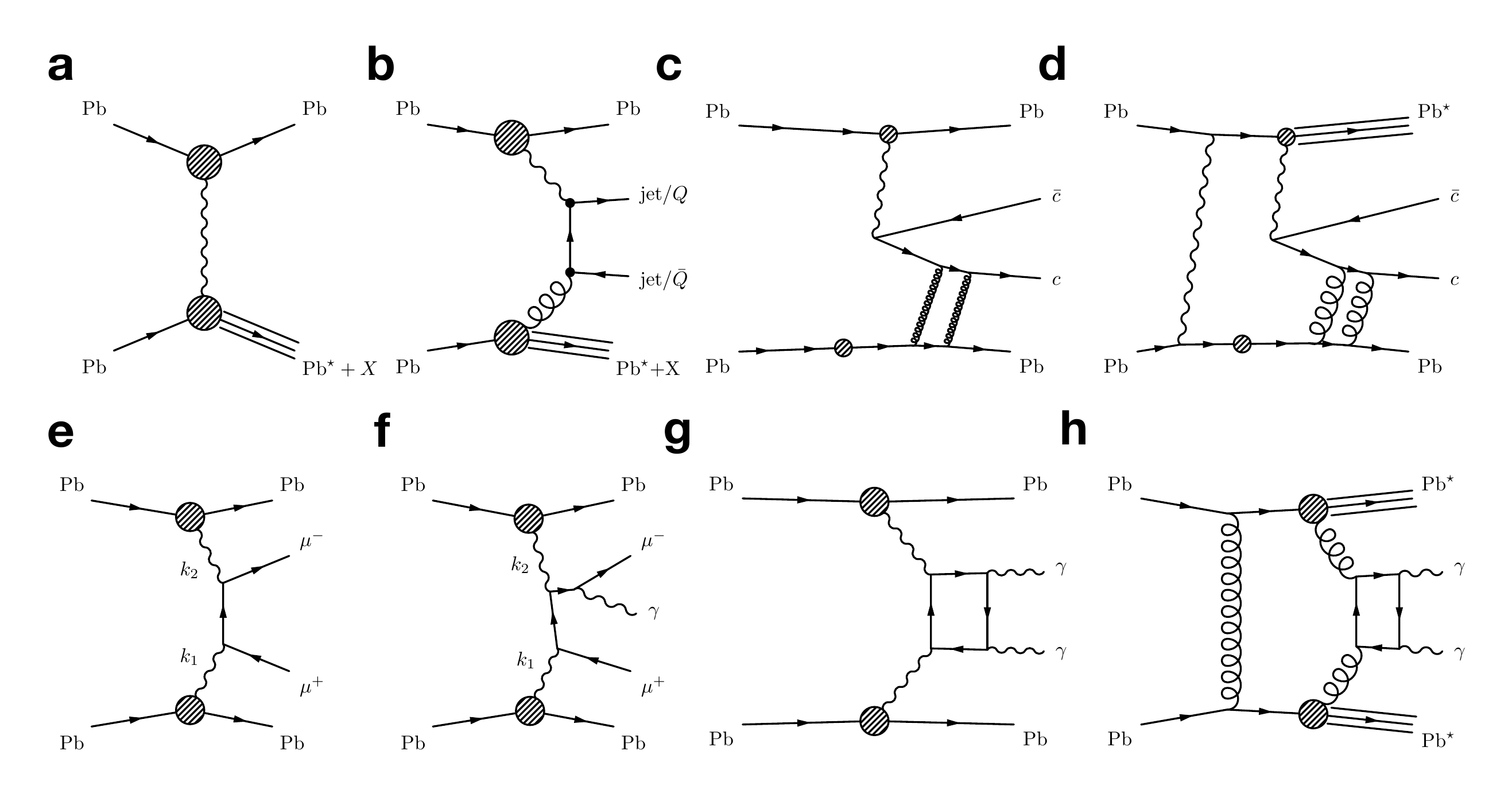}
\caption{Some of the UPC reactions that will be discussed in this review:
(a) generic photonuclear interaction with neutron breakup of the target, 
(b) incoherent photoproduction, generic to heavy quarks and jets,
(c) exclusive photoproduction of a vector meson 
(d) coherent photoproduction of a vector meson, accompanied by nuclear excitation, 
(e) dilepton production $\gamma\gamma\rightarrow l^+l^-$
(f) dilepton production $\gamma\gamma\rightarrow l^+l^-+\gamma$, including higher order final-state radiation
(g) light-by-light scattering, with no nuclear breakup
(h) central exclusive diphoton production, with double breakup.}
\label{fig:Feynman}
\end{figure}

Ultra-peripheral collisions (UPCs) involve collisions of relativistic nuclei (heavy ions or protons)  %SKXthat occur 
at impact parameters ($b$) that are large enough so that there are no hadronic interactions.  Instead, the ions interact electromagnetically, either via photonuclear or two-photon interactions.  In UPCs, the photons are nearly real, with virtuality $Q^2<(\hbar/R_A)^2$, where $R_A$ is the nuclear radius.   Typical photonuclear
interactions are vector meson photoproduction or production of dijets.   Typical $\gamma\gamma$ interactions lead to final states such as dileptons, single mesons or meson pairs, or two photons (via light-by-light scattering).  
Figure \ref{fig:Feynman} shows some of the reactions that will be discussed here.  Ultra-peripheral collisions have previously been reviewed elsewhere \cite{Bertulani:1987tz,Baur:2001jj,Bertulani:2005ru,Baltz:2007kq,Contreras:2015dqa,Klein:2017nqo}; the focus here is on newer developments.   We will also briefly discuss photoproduction and two-photon interactions in peripheral hadronic collisions.

This review will primarily focus on collisions involving nuclei and/or protons at Brookhaven's Relativistic Heavy Ion Collider (RHIC) and at CERN's Large Hadron Collider (LHC).
The LHC collisions are the energy frontier for photonuclear and two-photon physics, while RHIC typically provides  higher integrated luminosities and photon energies well suited for photonuclear interactions involving Reggeon exchange.  We also briefly consider collisions at CERN's proposed future circular collider (FCC), China's proposed SPPC \cite{CEPC-SPPCStudyGroup:2015csa},
and AFTER, a proposed fixed-target experiment utilizing a beam extracted from the LHC \cite{Massacrier:2017lib}. AFTER has a lower maximum $\gamma p$ center of mass energy, $W_{\gamma p}\approx 2$ GeV/c, but offers higher luminosity.   Similar physics is accessible in principle at electron-ion (including proton) colliders such as EIC, LHeC and FCC-eh, while two-photon physics can also be studied at $e^+e^-$ colliders.  
Table 1 gives the maximum energies for different ion species at these machines. Nuclear beams provide several distinct advantages

\begin{enumerate}
    \item a large effective photon luminosity boost proportional to $Z^2$ for each nucleus, compensating for the overall lower luminosity of nuclear beams
    \item reduced virtuality
    \item the possibility of multi-photon exchange between a single ion pair, allowing for tagging of different impact parameter distributions and photon spectra.
\end{enumerate}

Early UPC studies largely focused on $e^+e^-$ pair production and low-energy nuclear physics \cite{Bertulani:1987tz}. In the late 1980's, interest grew in using UPCs to probe fundamental physics, most notably two-photon production of the Higgs \cite{Papageorgiu:1988yg,Cahn:1990jk}.  Although the resulting $\gamma\gamma$ luminosities were not encouraging for observing the Higgs, they did stimulate work on $\gamma\gamma$ production of other particles.  The first calculations of coherent photoproduction with gold beams at RHIC predicted high rates of vector meson photoproduction \cite{Klein:1999qj}, which were quickly confirmed by the STAR Collaboration \cite{Adler:2002sc}.  The combination of large cross-sections and available experimental data stimulated further interest.  With the advent of the LHC, the energy reach for UPCs extended dramatically,
and the field has blossomed.  

% short MC section
A key to development of UPC as a precision laboratory for electromagnetic and strong interaction processes is the development
of event generators that simulate both the initial photon flux and the relevant physics processes.  The most widely-used generator code is $\SL$~\cite{Klein:2016yzr} which 
has been available since the early days of the RHIC program.  It implements one and two photon processes, and includes a set of final states including vector mesons, meson pairs, and dileptons, with more general photonuclear processes accessible using the DPMJET3 code.  
The SuperCHIC 3 Monte Carlo~\cite{Harland-Lang:2018iur} also implements nuclear photon fluxes, and computes many of the same processes.  Finally, event generators more commonly-used for proton-proton reactions, particularly $\PY$8~\cite{Helenius:2018mhx}, are getting more complete implementations of the coherent nuclear fluxes that provide opportunities to calculate final states needing parton showers, e.g. QCD jets.

\begin{textbox}
TABLE 1:
This table compares the capabilities of different colliders, in terms of available energy for photonuclear and photon-photon processes.
The table shows the accelerator, ion species, $\sqrt{s_{NN}}$, the maximum photon energy (in the target rest frame, relevant for cosmic-ray studies), $W_{\gamma p}$ (the $\gamma p$ photonuclear center-of-mass energy), and the maximum $\gamma\gamma$ energy for two-photon interactions.  For the photonuclear interactions, the maximum energies correspond to $x=1$ in Eq. \ref{eq:nofb} with $b=R_1+R_2$, the sum of the nuclear radii. For $\gamma\gamma$ interactions, the maximum photon energy is given for $x=1$ when $b$ is the nuclear radius. For the proton, we use $r=0.7$ fm; this is similar to treating the proton with a dipole form factor \cite{Klein:2003vd}.  This allows photons to carry up to about 30\% of the proton's energy - a higher cutoff than was used in some previous works \cite{Bertulani:2005ru}. For pA collisions, the maximum $W_{\gamma p}$ is higher when the photons  from proton strike the heavy ion than for the reverse case.  The table lists the opposite case, since the reaction rate for that direction is usually much higher.  For pPb at the LHC, the proton and ion beams have different Lorentz boosts, so the per-nucleon center of mass is boosted with respect to the lab frame. For the electron-ion colliders, we used the energies from the cited references (the designs are still evolving, so the beam energies may change), with the same maximum photon energy criteria as for UPCs.

\begin{center}
\begin{tabular}{|l|l|c|r|r|r|}
\hline
Facility & System & $\sqrt{s_{NN}}$ or $\sqrt{s_{eN}}$ &
Max. $E_{\gamma}$ & Max. $W_{\gamma p}$ & Max $\sqrt{s_{\gamma\gamma}}$ \\
\hline
RHIC & AuAu & 200 GeV & 320 GeV & 25 GeV & 6 GeV \\
     & pAu  & 200 GeV & 1.5 TeV & 52 GeV & 30 GeV \\
     & pp   & 500 GeV & 20  TeV &  200 GeV & 150 GeV \\
\hline
LHC \cite{Citron:2018lsq}
     & PbPb & 5.1 TeV  & 250 TeV & 700 GeV & 170 GeV \\
    & pPb  & 8.16 TeV & 1.1 PeV  & 1.5 TeV & 840 GeV \\
    & pp   & 14 TeV   & 16 PeV    & 5.4 TeV & 4.2 TeV \\
\hline
FCC-hh \cite{Benedikt:2018csr}
    & PbPb & 40 TeV & 13 PeV & 4.9 TeV & 1.2 TeV \\
 SPPC \cite{CEPC-SPPCStudyGroup:2015csa}    & pPb  & 57 TeV & 58 PeV & 10 TeV & 6.0 TeV \\ 
    & pp   & 100 TeV& 800 PeV & 39 TeV & 30 TeV  \\
\hline
\hline
eRHIC \cite{Montag:2019toh}
%SK the final energy is 110 GeV Au on 18 GeV e- %SK I had to change the reference, because the original one only considered operation at a lower electron energy - 10 GeV
     & eAu & 89 GeV & 4.0 TeV & 89 GeV & 15 GeV \\
\hline
%JLEIC removed after EIC site selection
%Updated with new, 2019 performance figures 80 GeV/n Au/Pb on 12 GeV e
%JLEIC \cite{JLEIC} 
%      & ePb & 62 GeV & 1.4 TeV & 44 GeV & 11 GeV \\
%\hline
% at least the LHeC energy hasn't changed much recently.
% this is 60 GeV electrons on 2.8 TeV/nucleon lead
LHeC \cite{AbelleiraFernandez:2012cc}
      & ePb & 820 GeV & 360 TeV  & 820 GeV & 146 GeV \\
\hline
\end{tabular}
\end{center}
\end{textbox}

\section{The photon flux from a relativistic ion}

\subsection{Impact parameter dependence}

A relativistic ion carries Lorentz-contracted electric and magnetic fields; the electric field radiates outward from the ion, while the magnetic field circles the ion. Fermi~\cite{Fermi:1924tc}, von Weizs\"acker~\cite{vonWeizsacker:1934nji} and Williams~\cite{Williams:1935dka} showed that these perpendicular fields may be treated as a flux of linearly polarized virtual photons; the energy spectrum is given by the Fourier transform of their spatial (along the ion direction) dependence. In the relativistic limit ($\beta\rightarrow 1)$ at a distance $b$ from an ion (where $b>R_A$), the photon energy ($k$) spectrum from an ion with charge $Z$, velocity $\beta c$, Lorentz boost $\gamma$ is
\begin{equation}
N(k,b)= \frac{Z^2\alpha k^2}{\pi^2\gamma^2\hbar^2\beta^2}\left(K_1^2(x)+\frac{K_0^2(x)}{\gamma^2}\right),
\label{eq:nofb}
\end{equation}
where $K_1$ and $K_0$ are Bessel functions and $x=kb/\beta\gamma\hbar c$.  For $x<1$, $N(k,b)\propto 1/x^2$, while, for $x>1$, the flux is exponentially suppressed.  The larger the photon energy, the smaller the range of $b$ that can contribute to the flux. 

\begin{marginnote}[]
\entry{Impact parameter}{$b_i$ is the generic distance from one of the ions, while $b_1$ and $b_2$  are the distances from each of the two ions. $b$ is the ion-ion impact parameter.}
\entry{Momentum}{$k$ is a photon momentum, $k_1$ and $k_2$ are the photon momenta in two-photon interactions, $q$ is for the exchange gluon/Pomeron, capital $P$ the momentum of the final state}
\entry{Momentum transfer}{$t$ is the squared momentum transfer, which is  $q^2$ in an elastic process.}
\entry{Pairs}{$M_{\ell \ell}$ is the final state mass for a di-lepton state, and $Y_{\ell \ell}$ is the final state pair rapidity.}
\end{marginnote}

The values of $\beta$ and $\gamma$ are frame-dependent. At colliders, the Lorentz boost of the photon-emitting nucleus in the target rest frame is $\Gamma=2\gamma^2-1$, so photon energies in the PeV range (1 PeV = $10^{15}$ eV) are accessible in the target rest frame.  The target frame is useful for comparison with cosmic-ray air showers.  

The total photon flux is found by integrating Eq. \ref{eq:nofb} over $b$.  The integration range depends on the application.  For ultra-peripheral collisions, we exclude collisions where the nuclei interact hadronically, to allow for reconstruction of exclusive final states. This can be done by taking the minimum impact parameter $b_{\rm min}$ to be $2R_A$.  The total flux is
\begin{equation}
N(k)= \frac{Z^2\alpha k^2}{\pi^2\gamma^2\hbar^2\beta^2}\left(K_1^2(u)+\frac{K_0^2(u)}{\gamma^2}\right),
\label{eq:nofb3}
\end{equation}
where $\alpha\approx 1/137$ is the fine structure constant and $u=\gamma\hbar c/b_{\rm min} = \gamma\hbar c/2R_A$.

Equation \ref{eq:nofb3} ignores the nuclear skin thickness (about 0.5 fm) and the range of the strong interaction.  The flux can be more accurately calculated with
\begin{equation}
N(k)= \int d^2b N(k,b) \Pnohad (b),
\label{eq:nofk}
\end{equation}
where $\Pnohad (b)$ is the probability of not having a hadronic interaction. $\Pnohad (b)$ can be calculated with a Glauber calculation \cite{Miller:2007ri} which accounts for the nuclear shape and interaction probability.  In these calculations, the nucleon distribution of heavy nuclei is well described by a Woods-Saxon distribution, while a Gaussian form factor is appropriate for lighter nuclei ($Z\le 6$) \cite{Klein:2016yzr}.  For protons, a dipole form factor is found to work well \cite{Drees:1988pp,Klein:2003vd,Klein:2018grn}.  This corresponds to an exponential charge distribution. 

\subsection{Nuclear dissociation}
For heavy nuclei, $Z\alpha\approx 0.6$, so the probability of exchanging more than one photon between the two ions in an individual collision must be considered. These photons are essentially independent of each other, even if they are emitted by the same nucleus \cite{Baur:2003ar}. The additional photons may dissociate one or both nuclei, or, less often, introduce additional particles into the detector.  Because of the radial dependence of the photon flux, the presence of these additional photons can preferentially select certain impact parameter ranges, and so can influence the photon spectrum of the other photons. 

Some triggers or analyses may require either the presence or absence of neutrons in forward zero-degree calorimeters (ZDCs). Additional photons may break up one or both nuclei, producing neutrons. Mutual Coulomb Excitation (MCE), via two additional exchanged photons generally leads to neutrons in both ZDCs.  

\begin{figure}[t]
\includegraphics[width=0.90\textwidth]{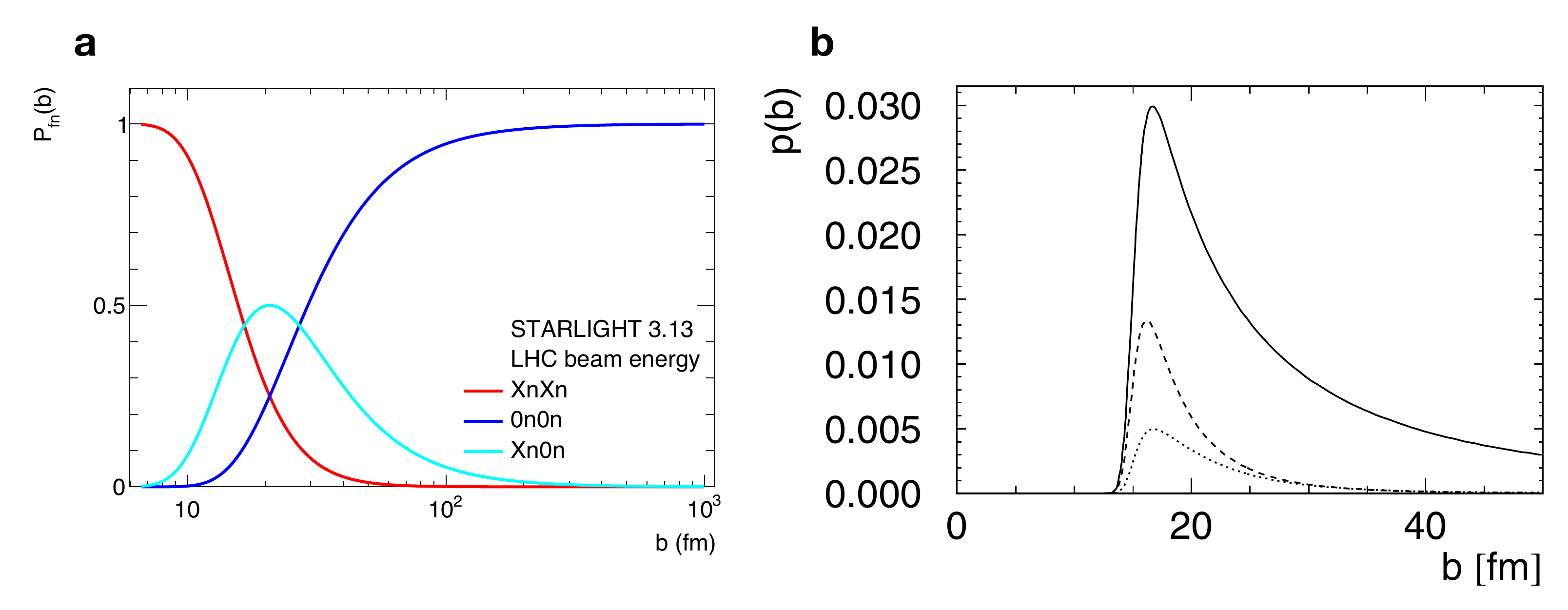}
\caption{(a.) Impact parameter dependence of the probabilities, from \SL\ 3.13, of the three primary forward neutron topologies: 1) $\ZOnOn$, or no neutron emission in either direction, which selects impact parameters $b>40$ fm 2) $\ZXnOn$, with neutron emission in only one direction, which selects impact parameters of $b\sim 20$ fm, and 3) $\ZXnXn$, with neutron emission in both directions, selecting impact parameters $b<15$ fm.
(b.) \SL\ calculation of the impact parameter dependence of coherent $\rho$ production, assuming different ZDC fragmentation scenarios, from Ref.~\cite{Baltz:2002pp}. The difference between 
MCE and no breakup is quite stark at larger impact parameters.}
\label{fig:starlight_lhc_breakup}
\end{figure}

If a harder photon spectrum is desired, one can require neutrons in one or both ZDCs to select events with additional Coulomb excitation, {\it i. e.} with smaller impact parameters.  Adding breakup conditions leads to
\begin{equation}
   N(k)= \int d^2b N(k,b) \Pnohad (b) P_1 (b) P_2(b),
\label{eq:nofk2} 
\end{equation}
where the $P_1$ and $P_2$ are the excitation probability for the two nuclei.  Generally, $P_{i}(b)\propto 1/b^2$, so requiring nuclear excitation leads to smaller impact parameters. 
This is demonstrated using \SL\  for the three possible cases in Fig~\ref{fig:starlight_lhc_breakup} (a), which shows the functional forms of $P_{fn}(b)$, the impact-parameter-dependent probability of a particular forward neutron configuration: $\ZXnXn$ ($P_1 (b) P_2(b)$), $\ZXnOn$ ($(1-P_1)P_2+(1-P_2)P_1$), and $\ZOnOn$ ($(1-P_1)(1-P_2)$).
Calculations from Ref.~\cite{Baltz:2002pp}, shown in Fig~\ref{fig:starlight_lhc_breakup} (b), demonstrate the stark differences in the impact parameter dependence of $\rho$ photoproduction for different ZDC topologies ($\ZXnXn$, $1n1n$, and $\ZOnOn$ here) for collisions at RHIC.

Eq. \ref{eq:nofk2} can also be applied to meson photoproduction. At the LHC, when $b=2R_A$, the probability of producing a $\rho^0$ is about 3\%; assuming that the electromagnetic fields are not depleted, the probability of $\rho$ photoproduction should be Poisson distributed, so the probability of producing two $\rho^0$ is about $5\times10^{-4}$. This is roughly 1 million pairs in a 1-month LHC heavy ion run \cite{Klein:1999qj}.  More exotic pairs, like $\rho J/\psi$, should also be visible.
These pairs are of interest because the two vector mesons should share a common polarization \cite{Baur:2003ar}.  Since they are bosons, there should be an enhanced probability to produce the two mesons in the same final state, and, potentially, observe stimulated decays. 

For some applications, the photon flux within the nucleus ($b<R_A)$ is of interest. Two examples are $\gamma\gamma$ production of lepton pairs, which may occur within one of the nuclei (without dissociating it), even when $b>2R_A$,  and for studying electromagnetic processes in peripheral collisions, where there are also hadronic interactions.  At transverse distance $b_i<R_A$ from the nuclear center, only nucleons in a cylinder with radius $b_i$ contribute coherently to the interaction  \cite{Zha:2017jch}.  This can be handled by adding a form factor to the photon emission flux~\cite{Vidovic:1992ik,Vidovic:1993cf,Zha:2018tlq}.  It is also possible that the hadronic interaction might disrupt the coherent photon emission.  However, since the photons are nearly real, they should be mostly emitted at a time before the hadronic interaction occurs.  

\subsection{$k_T$ spectrum}
\label{subsec:pt}

The photon $k_T$ spectrum can be derived from the equivalent photon approximation (EPA).  If one integrates the photon flux over all $b$, then the photon $k_T$ may be determined exactly 
\cite{Vidovic:1992ik}
\begin{equation}
    \frac{d^2N}{d^2k_{T}} = \frac{\alpha Z^2 F(k_T^2+k_z^2/\gamma^2) k_T^2}{\pi^2 (k_T^2+k_z^2)^2},
\end{equation}
where $F(k^2)$ is the nuclear form factor.  When integrated over $k_T$, this gives Eq. \ref{eq:nofk}.  

If the range of $b$ is restricted, such as by requiring $b>2R_A$ 
%SKZ(or $b>R_A$) 
or by weighting the $b$-distribution by the requirement that there be an additional photon or photons exchanged, the problem becomes much more complicated, because $k_T$ and $b$ are conjugate variables.   
This is seen in more complete QED calculations, e.g. Ref.~\cite{Vidovic:1992ik} and ~\cite{Zha:2018tlq}. As the range of $b$ is restricted, the mean $k_T$ should increase.  Unfortunately, we do not yet have a method to calculate the $k_T$ spectrum for these cases.  
Calculations for two-photon interactions are more complicated because the $\gamma\gamma$ interaction point is distinct from the radial positions relative to each of the two nuclei. 
In principle, restrictions on the ion-ion impact parameter do not affect the photon $p_T$ spectrum at this third point. 
However, final states clearly identified with $\gamma\gamma\rightarrow l^+l^-$ show significant $k_T$ broadening as the impact parameter range is restricted to smaller values in hadronic heavy ion collisions \cite{Adam:2018tdm,Aaboud:2018eph,ATLAS-CONF-2019-051}, as we discuss below.

\subsection{Uncertainties on the photon flux}

The photon flux calculations are subject to a number of theoretical uncertainties:
\begin{enumerate}
    \item  {\bf Overlap condition} The implementation of $\Pnohad$ is straightforward, but implemented using optical Glauber calculations, which are known to be limited in applicability.
    \item {\bf Restriction on production points} It is typically assumed that UPC processes can not originate within a nucleus, i.e. $b_i>R_i$, but this is not clearly established using data.      It is possible that some range within a nucleus could be accessible to these processes, contributing to the observed enhancement of $\Jpsi$ in peripheral collisions.  
    \item {\bf Assumption of a uniform flux} Normally, the photon flux is taken to be constant across the entire target nucleus, well represented by a plane wave.  This assumption ignores the fact that the maximum photon energy (and flux) %SKZ have a strong dependence  
    are higher on the near side of the nucleus, and lower on the far side.  
\end{enumerate}

For two-photon interactions, the effect of changing impact parameter cutoffs has been studied.  The uncertainty rises with increasing $W_{\gamma\gamma}/\sqrt{s_{nn}}$, reaching about 5\% at half of the maximum $W_{\gamma\gamma}$ in Table 1 \cite{Nystrand:1998hw}.
For the others, precise experimental measurements are required to assess their relative importance.

\section{Photoproduction processes: $\gamma+p$ and $\gamma+A$}

\subsection{Low energy photonuclear interactions (including nuclear dissociation)}

Since the photon spectrum scales roughly as $1/k$, the most common photonuclear interactions involve low-energy nuclear excitations.   The Coulomb excitation with the largest cross-section is the Giant Dipole Resonance (GDR).  In it, protons and neutrons oscillate collectively, against each other \cite{Berman:1975tt}. The GDR has the same quantum numbers as the photon $J^{PC}=1^{--}$, so it is readily excited by them.  GDRs decay primarily by single neutron emission, while most higher excitations involve the emission of multiple neutrons.  For this reason, it is a useful calibration signal.  The total excitation cross-section is determined by combining the photon spectrum with the photonuclear excitation cross-section:
\begin{equation}
\sigma({\rm Excitation}) = \int dk \frac{dN}{dk}\sigma(\gamma A \rightarrow A^\star).
\label{eq:photoexcite}
\end{equation}
$\sigma(\gamma A \rightarrow A^\star)$ is determined from a compilation of photoexcitation data \cite{Baltz:1996as}, or, in some cases, from first principles~\cite{Pshenichnov:2001qd}.  Because these cross-sections can be very large, corrections may be needed to account for unitarity, because without them, $P_i(b)$ can be large for $b\approx 2R_A$  \cite{Baltz:1996as}.  Multiple photon absorption leads to higher excitations.  The Coulomb excitation cross-section is about 95 barns with gold-gold collisions at RHIC, and 220 barns with lead-lead collisions at the LHC .

Interactions with different requirements on nuclear breakup (or non-breakup, where we require no observed neutrons) can be calculated in an impact parameter dependent formalism.  For example, MCE primarily occurs via two-photon exchange, with each photon exciting one nucleus.  The cross-section for MCE is
\begin{equation}
\sigma(\ZXnXn) = \int d^2b P_1(b) P_2(b) P_{0\mathrm{}{had}} (b),
\end{equation}
where the respective probabilities are for exciting nucleus 1, exciting nucleus 2, and not having a hadronic interaction. The excitation probabilities can be determined from Eq. \ref{eq:photoexcite}, using the impact-parameter-dependent photon flux.  

MCE can be used to monitor luminosity.  The most delicate part is accurately determining $\Pnohad(b)$ when $b\approx 2R_A$.  This uncertainty can be avoided by instead calculating and measuring the summed cross-section for MCE plus hadronic interactions \cite{Baltz:1998ex}.  Alternately, events with one neutron in each ZDC generally correspond to mutual GDR excitation which has lower backgrounds from hadronic interaction than MCE in general\cite{Adamczyk:2017vfu}, although the cross-sections are lower, so the statistics are limited.

The neutron multiplicity distribution in Coulomb exchange processes has been studied by several groups \cite{Chiu:2001ij}, most recently by the ALICE experiment \cite{ALICE:2012aa}, whose ZDCs could separate events containing from one to four neutrons.  Their measurements were in generally good agreement with the predictions of the RELDIS model \cite{Pshenichnov:2001qd}, which uses the Weizsacker-Williams photon flux, measured photonuclear cross-sections and neutron emission via cascade and evaporation codes.  The n$^0_0$n afterburner \cite{Broz:2019kpl} performs similar calculations, but in a Monte Carlo format which can be used with existing simulations to simulate vector meson photoproduction with nuclear breakup.

\subsection{Probing nuclear parton distributions with incoherent photoproduction}

As Fig.~\ref{fig:Feynman}(b) shows, the photon-gluon fusion process directly probes
the gluons in the oncoming nucleus, so it can be used to directly study nuclear shadowing \cite{Klein:2019qfb}.   Triggering on such processes is aided by the fact that the photon emitter typically does not break up.  However the recoiling partons will in general excite the nucleus, leading to nuclear dissociation, and an accompanying partonic connection (`string') between the reaction products and the nuclear remnants. 

\begin{figure}[t]
\includegraphics[width=.97\textwidth]{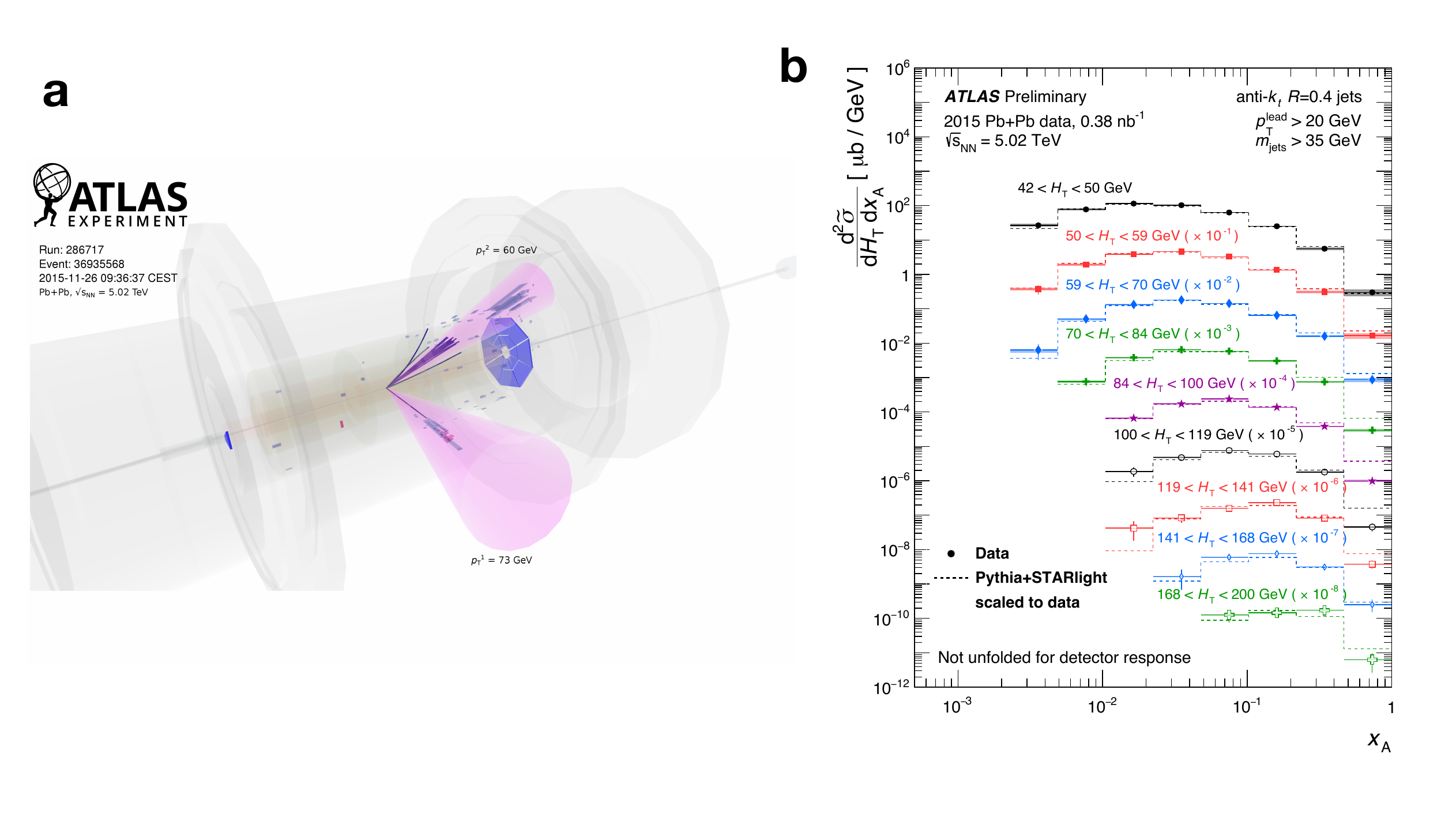}
\caption{(a.) Event display of an event with a large gap in one direction, with two jets in the other direction. (b.) Uncorrected triple differential cross sections, from Ref.~\cite{ATLAS-CONF-2017-011}.}
\label{atlas-dijets}
\end{figure}

This leads to a distinct event signature, with limited transverse energy, neutrons in only one direction, and one or more reconstructed jets, easily implemented in an experimental trigger. Fig.~\ref{atlas-dijets} (a) shows an example event, triggered using an exclusive one-arm ZDC trigger and containing two forward reconstructed jets.

Jets are straightforward to reconstruct, particularly in  low-multiplicity photonuclear events, but the condition that they are well-reconstructed imposes a minimum $\pT$, which restricts the kinematic coverage in Bjorken-$x$ (the fraction of the nucleon momentum) and $Q^2$, the hardness scale of the interaction.
ATLAS has presented preliminary triple-differential quasi-cross sections corrected for trigger efficiency, but not yet unfolded for experimental resolution~\cite{ATLAS-CONF-2017-011}.  The experimental variables are $H_T$, the scalar sum of the observed jet transverse momenta, $z_\gamma$, which reflects the energy fraction of the incoming nucleon energy carried by the photon, and $x_A$, the per-nucleon momentum fraction in the struck nucleus. The data are compared to a $\PY 6$ calculation, calculated using $\mu+N$ interactions, mediated by a nearly-real photon, with the photon spectrum reweighted to match \SL\, and finally with the normalization adjusted to match to the data.  The particular result shown in Fig.~\ref{atlas-dijets} (b) shows cross sections vs. $x_A$ for selections in $H_T$.  Although the data has not yet been compared in detail with $\PY$, the general agreement is very promising.

Photoproduction of open charm also proceeds via photon-gluon fusion.  Open charm can be detected relatively near threshold ($M_{cc}\approx {\rm few}\ m_c$), 
so it can probe lower $x$ and $Q^2$ gluons than dijets \cite{Adeluyi:2012ph}. Charmed quarks, with charge $+2/3e$ have a particularly large cross-section \cite{Klein:2002wm,Goncalves:2017zdx}.  The disadvantage is that charm quarks can hadronize to several different hadrons, each with many possible final states. Many of the final states are difficult to reconstruct, so the overall reconstruction efficiency is small. 
Open bottom should also be visible at rates high enough for parton distribution studies \cite{Strikman:2005yv}, and top quark pair production may be accessible in $pA$ collisions at the LHC \cite{Klein:2000dk,Goncalves:2013oga}.    

Inelastic photoproduction is also a topic of interest in its own right.  Proceeding by the fluctuation of the photon to a hadronic state, typically a $\rho$ meson, it provides another example of collective behavior in small systems.  Early results from ATLAS~\cite{ATLAS:2019gsn} indicate that these collisions show a "ridge" structure in the two-particle azimuthal correlation function, similar to that seen in $pp$, but with a smaller magnitude, possibly reflecting a more compact quark-antiquark configuration compared to the proton with its three constituent quarks.

\subsection{Coherent and incoherent photoproduction of vector mesons}

\subsubsection{Coherent Photoproduction}

Coherent photoproduction refers to reactions where the incident nuclei primarily remain intact, so the  final state consists of the two incident nuclei plus a vector meson.  This reaction does not transfer color, so it must proceed via the exchange of at least two gluons.  At high energies, this two gluon exchange is often referred to as `Pomeron exchange.'  The Pomeron has the same quantum numbers as the vacuum, $J^{PC}=0^{++}$, so it can also be described as representing the absorptive part of the cross-section \cite{Forshaw:1997dc}. 
At lower photon energies, meson photoproduction can also proceed via `Reggeon exchange,' where the Reggeons represent collective meson trajectories.  Reggeons carry a much wider range of quantum numbers than the Pomeron, and can be either neutral or charged.  So, the range of final state spin and parity is much wider, and charged final states and exotica are possible.  In UPCs, lower photon energies corresponds to production at forward rapidities, so this physics can best be studied with forward spectrometers.

The vector-meson rapidity distribution $d\sigma/dy$ can be converted to the incoming lab-frame photon and Pomeron energy via the relationships
\begin{equation}
k_{1,2}=\frac{M_V}{2}e^{\pm y}, \ \ \ q_{1,2}=\frac{M_V}{2}e^{\mp y},
\label{eq:photonenergy}
\end{equation}
where $M_V$ is the vector meson mass; we neglect the $k_T$ and $q_T$.  The $\pm$ and $\mp$ signs arise from the two-fold ambiguity over which nucleus emitted the photon. Away from $y=0$, the two possibilities have different photon and Pomeron energies.  This degeneracy complicates extraction of energy-dependent photoproduction cross-sections.  It can be largely avoided in $pA$ or other asymmetric collisions, where the photon comes predominantly from the heavy ion, with the proton as a target \cite{Sirunyan:2018sav,Sirunyan:2019nog,Acharya:2018jua}.  Unfortunately, this does not allow us to probe ion targets.  For pp collisions, HERA data can be used to fix the cross-section for one of the photon directions, allowing the cross-section for the other direction to be found \cite{LHCb:2016oce}.  For ion-ion collisions, another strategy is required: selecting sets of events with different photon spectra, leading to different ratios for production in the two directions. This can be done by selecting set of events with different impact parameter distributions.  For example, events accompanied by MCE have a harder photon energy spectrum \cite{Baltz:2002pp,Baur:2003ar}.  Or, one can compare photoproduction in peripheral and ultra-peripheral collisions \cite{Contreras:2016pkc}.  By using multiple data sets with different impact-parameter distributions, it is possible to unambiguously find the energy dependence of the cross-section $\sigma(k)$, albeit with increased errors due to the coupled equations.   

\subsubsection{Cross-section for $\gamma p$ interactions}

The bidirectional ambiguity problem goes beyond summing the cross-sections for the two photon directions. For coherent photoproduction, the reactions with different photon directions are indistinguishable, so they interfere with each other \cite{Klein:1999gv}.  The sign of the interference depend on the how the two possibilities are related.  A reaction where ion 1 emits a photon can be transformed into a reaction where ion 2 emits a photon by a parity exchange.  Vector mesons are negative parity, so the interference is destructive.  For $\overline pp$ collision at the Fermilab Tevatron, the relevant symmetry is $CP$, so the interference is constructive.  There is also a propagator $\exp{(i\vec{p}_T\cdot \vec{b})}$ present.  For an interaction at a given impact parameter
\begin{equation}
\frac{d^2\sigma}{dbdy} = \big|A_1(b,k_1) - e^{i\vec{p}_T\cdot \vec{b}} A_2(b,k_2)\big|^2,
\label{eq:interfere}
\end{equation}
where $A_1(b,k_1)$ and $A_2(b,k_2)$ are the amplitudes for the two directions.   When $y\approx 0$, in the limit $\pT\rightarrow 0$, interference is complete, and $\sigma\rightarrow 0$.  Away from $y=0$, the photon energies corresponding to the two directions are different, so $A_1(b,k_1)$ and $A_2(b,k_2)$ are different, reducing the degree of interference.   When $\pT\gg \hbar/R_A$, the propagator oscillates rapidly with small changes in $b$, and so averages out. Because of the oscillatory behavior, this interference does not significantly affect the total cross-section.
As the inset in Fig. \ref{fig:cms-alice-jpsi} (b) shows, suppression is visible for $\pT < 30$ MeV \cite{Adamczyk:2017vfu}, at the expected level \cite{Abelev:2008ew}.  

One interesting aspect of this interference is that the production amplitudes at the two nuclei are physically separated and share no common history.  Moreover, the two vector meson amplitudes decay almost immediately.  The $\rho^0$ has lifetime  of order $10^{-23}$ s, producing two pions which travel in opposite directions.  These pions become physically separated long before the wave functions for production at the two ions can overlap.  With the interference-imposed requirement that the pair $\pT$ not be zero, the pion pair can only be described with a non-local wave function.  This is an example of the Einstein-Podolsky-Rosen paradox \cite{Klein:2002gc}. 

$\gamma p$ interactions have been extensively studied at fixed-target experiments \cite{Bauer:1977iq} and the HERA $ep$ collider \cite{Abramowicz:1998ii}. A wide range of final states have been observed.  At high energies, Pomeron exchange dominates, and the most common final states are vector mesons, including the $\rho$, $\omega$, $\phi$, $J/\psi$, $\psi'$ and $\Upsilon$ states.  Direct $\pi^+\pi^-$ pairs are also produced.  Their production may be modelled as photon fluctuations directly to a pair of charged pions. 

The cross-section to produce a vector meson $V$ depends on the probability for the photon to fluctuate to a $q\overline q$ pair and on the cross-section for that pair to scatter elastically from the target, emerging as a real vector meson. The fluctuation probability depends on the quark charge and vector meson wave function; it is quantified with the coupling $f_v$, which is determined from $\Gamma_{ll}$, the leptonic partial width for for $V\rightarrow e^+e^-$.  The elastic scattering cross-sections are then determined using $\sigma(\gamma p\rightarrow Vp)$  and $f_v$.

In lowest order perturbative QCD, the reaction proceeds via two gluon exchange.  Two gluons are required to preserve color neutrality.
The forward scattering cross-section to produce a vector meson with mass $M_V$ is often given as \cite{Jones:2013pga}
\begin{equation}
\frac{d\sigma}{dt}\bigg|_{t=0} = \frac{\Gamma_{ll}M_v^3\pi^3}{48\alpha}
\bigg[\frac{\alpha_S({\overline Q}^2)}{\overline Q^4}xg(x,\overline Q^2\big)^2\big(\frac{1+Q^2}{M_V^2}\big)\bigg],
\label{eq:lopqcd}
\end{equation}
where 
% repeat $\Gamma_{ll}$ is the partial width for $V\rightarrow e^+e^-$,
$\alpha_S$ is the strong coupling constant, $x$ is the Bjorken$-x$ of the gluon and  $\overline{Q}^2=(Q^2+M_V^2)/4$, where $Q$ is the photon virtuality, which is generally small.  The division by four is because there are two gluons, each assumed to carry half the virtuality; as we will see later, this assumption is problematic. The vector meson mass provides a hard scale, allowing the use of perturbative QCD, even when the photon virtuality is small.  pQCD is usually assumed to be applicable for photoproduction of the $J/\psi$ and heavier mesons.  Two gluons form the simplest color neutral object that can be exchanged.  More sophisticated models treat Pomeron exchange as a gluon ladder \cite{Collins:1992cv}. 

The two-gluon approach has some important caveats  \cite{Klein:2017vua}, many of which also apply to incoherent photoproduction.  The gluon density, $g(x,\overline Q^2)$ is squared, to account for the two gluons, but there is no reason the two gluons should have the same $x$ and $Q^2$ values.  In fact, the largest contribution occurs when the two gluons have very different $x$ values,  $x_1\gg x_2$, so the softer gluon is relatively unimportant.  One way to account for the soft gluon is to treat the interaction using a generalized parton distribution (GPD).   Another approach is to account for the second gluon using a Shuvaev transformation \cite{Shuvaev:1999ce}; this leads to a multiplicative factor in Eq. \ref{eq:lopqcd}.  As long as $x_1\gg x_2$, the Bjorken$-x$ of the dominant gluon can be related to the Pomeron energy from Eq. \ref{eq:photonenergy} via $x=q/m_p = M_V/m_p \exp{(\mp y)}$, where $m_p$ is the proton mass.

There are also small corrections to account for higher order photon fluctuations (`resolved photons'), such as to $q\overline q g$.   Finally, there is some uncertainty due to the choice of mass scale, $\mu=\overline{Q}$, used to evaluate the gluon distribution \cite{Jones:2013pga}.  All of these considerations are the subject of intense theoretical discussion \cite{Flett:2019ept}.  One important next step is to extend the calculation to next-to-leading order (NLO).  There is not yet a complete NLO calculation, but most of the elements exist, and there are several partial-NLO results.

The NLO calculation includes contributions from many Feynman diagrams, including several where the quark distributions in the target are important.  A problem arises with the NLO calculation, at least at LHC energies: the NLO amplitude is larger than the LO amplitude.  This is because of the parton distributions used as input.  There is no gluon data in this $x,Q^2$ range, so the parton distributions extrapolate downward in $x$, finding a very small gluon contribution.  This leads to a very small LO cross-section, so it is unsurprising that the NLO diagrams give a larger contribution.  By starting with a larger gluon contribution, such as can be inferred from the $J/\psi$ coherent photoproduction data, this problem disappears.

The cross-section is sensitive to the choice of factorization and renormalization scales and the choice of the minimum virtuality to consider. These can be reduced with a careful choice of minimum virtuality, $Q_0$ \cite{Jones:2016ldq}.  With an optimal choice, the scale uncertainty is reduced to the $\pm 15\%$ to $\pm 25\%$ range for the $\Upsilon$, somewhat larger for the $J/\psi$.  This range is not small, but it is small enough to allow for photoproduction to meaningfully contribute to parton distribution fits in this $x,Q^2$ range, even with the systematic uncertainties \cite{Flett:2019pux}. 

The longitudinal momentum transfer $q_z$ in photoproduction is determined by the kinematics: $q_z=4k/M_V^2$.  The maximum momentum transfer is set by the coherence condition applied to the proton size $R_p$, $q<\hbar/R_p$.  More precisely, the $\pT$ is regulated by the proton form factor, and the cross-section can be written
\begin{equation}
\sigma(\gamma p\rightarrow Vp) = \frac{d\sigma}{dt}\bigg|_{t=0} \int_{t_{\rm min}}^\infty  |F(t)|^2 dt,
\label{eq:forwardVM}
\end{equation}
where $t\approx p_T^2$ and $F(t)$ is the proton form factor and $t_{\rm min}$ is the minimum momentum transfer, $M_V^2/2k$ (with $k$ in the target frame).
$d\sigma/dt|_{t=0}$ encodes all of the hadronic physics of the reaction.   Because Eq. \ref{eq:lopqcd} is decoupled from the nuclear form factor in  Eq. \ref{eq:forwardVM}, in this approach, changes in the gluon distribution do not alter the nuclear shape. 

Several groups have used UPCs to study $J/\psi$ photoproduction on proton targets, extending the data collected at HERA.  Above the threshold region, HERA and fixed-target experiments found that the $J/\psi$ photoproduction cross-section is well described by a power law $\sigma\propto W_{\gamma p}^\alpha$, with $\alpha =0.67\pm0.03$ \cite{LHCb:2016oce}.  This linear relationship is expected in LO pQCD, Eq. \ref{eq:lopqcd} as long as the low$-x$ gluon distribution itself follows a power-law, $g(x,Q^2)\propto x^{-\alpha/2}$.  A deviation from this power law would signal that higher order diagrams are becoming important, a possible precursor to saturation.  The ALICE collaboration used pPb collisions to extend the measurement up to roughly $W_{\gamma p}$=800 GeV, probing gluons with $x\approx 2\times 10^{-5}$ and finding no deviation from the power-law behavior \cite{Adam:2016lmr}.  In the overlap region, the data were in good agreement with the HERA data.  It should be noted that, at these energies NLO calculations predict $J/\psi$ cross-sections similar to LO. 

The LHCb collaboration made a similar study in pp collisions at $\sqrt{s}$=7 TeV and $\sqrt{s}$=13 TeV \cite{LHCb:2016oce}.  They used HERA data to fix the cross-section in the direction corresponding to the low-energy photon solution, and then solved for the cross-section in the high-energy photon direction.  At $\sqrt{s}$=13 TeV, their highest rapidity point, $\langle y\rangle=4.37$ corresponded to $W_{\gamma p} \approx 1.7$ TeV and $x\approx 3\times 10^{-6}$.  Unfortunately, the results from the different energies show some tension. The $\sqrt{s}$=7 TeV data follows the HERA power law, while the $\sqrt{s}$=13 TeV is lower, consistent with a NLO prediction.  The found similar power law behavior for the $\psi'$, albeit with larger statistical uncertainty \cite{McNulty:2016sor}. 

The $\Upsilon$ states are of interest because the LO and NLO calculations differ more especially with increasing collision energy,  making them more sensitive to the presence of higher order contributions \cite{Jones:2013pga}. 
LHCb has also observed photoproduction of the three $\Upsilon$ states \cite{McNulty:2016sor}.  For the $\Upsilon(1S)$, where the statistics are best, they observe good agreement with their NLO calculations, above the LO predictions.   

The CMS collaboration studied $\rho$ photoproduction on protons in pPb collisions, focusing on the $p_T$ spectrum.  They showed a pion pair $p_T$ spectrum out to 1 GeV/c \cite{Sirunyan:2019nog} and found that it was well described by a mixture of exclusive interactions, plus `incoherent' interactions, where the proton dissociated, and $\rho^0 (770)$ feeddown from $\rho(1700)$ decays.  From this, they extracted the $\rho(770)$ component of the dipion spectrum, and found that the cross-section was in agreement with HERA predictions, but that $d\sigma/dt$ dropped faster than an exponential. Their data was well fit by the form $\exp(-bt+ct^2)$, with $c\approx 3\sigma$ from zero. This indicates that the proton size depends on the $Q^2$ at which it is observed. 

CMS used a similar technique to study the $\Upsilon$ in pPb collisions \cite{Sirunyan:2018sav}.  They resolved the $\Upsilon(1S)$ and$\Upsilon (2S)$ peaks, and concentrated on the $\Upsilon(1S)$, where the statistics were better.  Because of the scaling with $Z$, the continuum background from $\gamma\gamma\rightarrow\mu\mu$ was substantially larger in these pPb collisions than in pp collisions.  The cross-section was consistent with a power law in $W_{\gamma p}$, at a level between the NLO and LO predictions shown by CMS.  

\subsubsection{Cross-section for $\gamma A$ interactions}

One driver for studies of vector meson photoproduction in $\gamma A$ is to study how the parton distributions in nucleons change when they are embedded in nuclei - the phenomenon of "nuclear shadowing".  Ignoring  shadowing, the cross-section may be found via a Glauber calculation \cite{Frankfurt:2002sv}:
\begin{equation}
\frac{d\sigma_{\gamma A\rightarrow VA}}{dt} =
\frac{d\sigma_{\gamma p\rightarrow Vp}}{dt}\big|_{t=0}
\ \bigg|\int d^2b\int dz e^{i(\vec{q_t}\cdot\vec{b}+q_{l}z)}
\rho(b,z)
e^{-\frac{1}{2}\sigma_{\rm tot}(Vp) T_{A}(b,z)}
\bigg|^2,
\label{eq:glauber}
\end{equation}
where 
\begin{equation}
T_A(z) = \int_z^\infty \rho(b,z')dz'
\end{equation}
and $\rho(b,z)$ is the nuclear density. $\sigma_{\rm tot}(Vp)$ is the total vector meson-nucleon cross-section, determined using the optical theorem:
\begin{equation}
\sigma^2_{\rm tot}(Vp) = 16\pi\frac{d\sigma (Vp\rightarrow Vp)}{dt}\big|_{t=0}.
\end{equation}
Here the elastic scattering cross-section $d\sigma(Vp\rightarrow Vp)/dt$ is determined from the measured photoproduction cross-section, after factoring out the $\gamma\rightarrow V$ fluctuation probability.   The exponential $\exp{(i(\vec{q_t}\cdot\vec{b}+q_{l}z))}$ accounts for coherence across the nucleus. 

The Glauber calculation accounts for multiple interactions - a single dipole encountering a nucleus may interact more than once, but can only produce a single vector meson.  For small
$\sigma_{\rm tot}(Vp)$ (i.e., heavy mesons like the $J/\psi$),  multiple interactions are unlikely, the amplitudes add linearly and the forward cross-section $d\sigma/dt|{t=0}$ scales as $A^2$.  Light mesons, with large $\sigma_{\rm tot}(Vp)$  will interact on the front surface of the target nucleus, so the amplitude depends on the frontal surface area of the target, $A^{2/3}$ and $d\sigma/dt|_{t=0}\propto A^{4/3}$.  The Glauber calculation accurately interpolates between these limits, under the assumption that the $\gamma p$ cross-section is the same for isolated protons, and for those in nuclei.   The range of $t$ for which coherent photoproduction is possible decreases as $A^{-2/3}$, moderating the increase in total coherent cross-section. If significant shadowing is present, then the cross-section is reduced below the Glauber expectation. For heavy nuclei at RHIC/LHC energies, the resulting cross-section is only slightly dependent on the photon energy, even if the $\gamma p$ cross-section shows a significant photon energy dependence.  

\begin{figure}[t]
\includegraphics[width=0.9\textwidth]{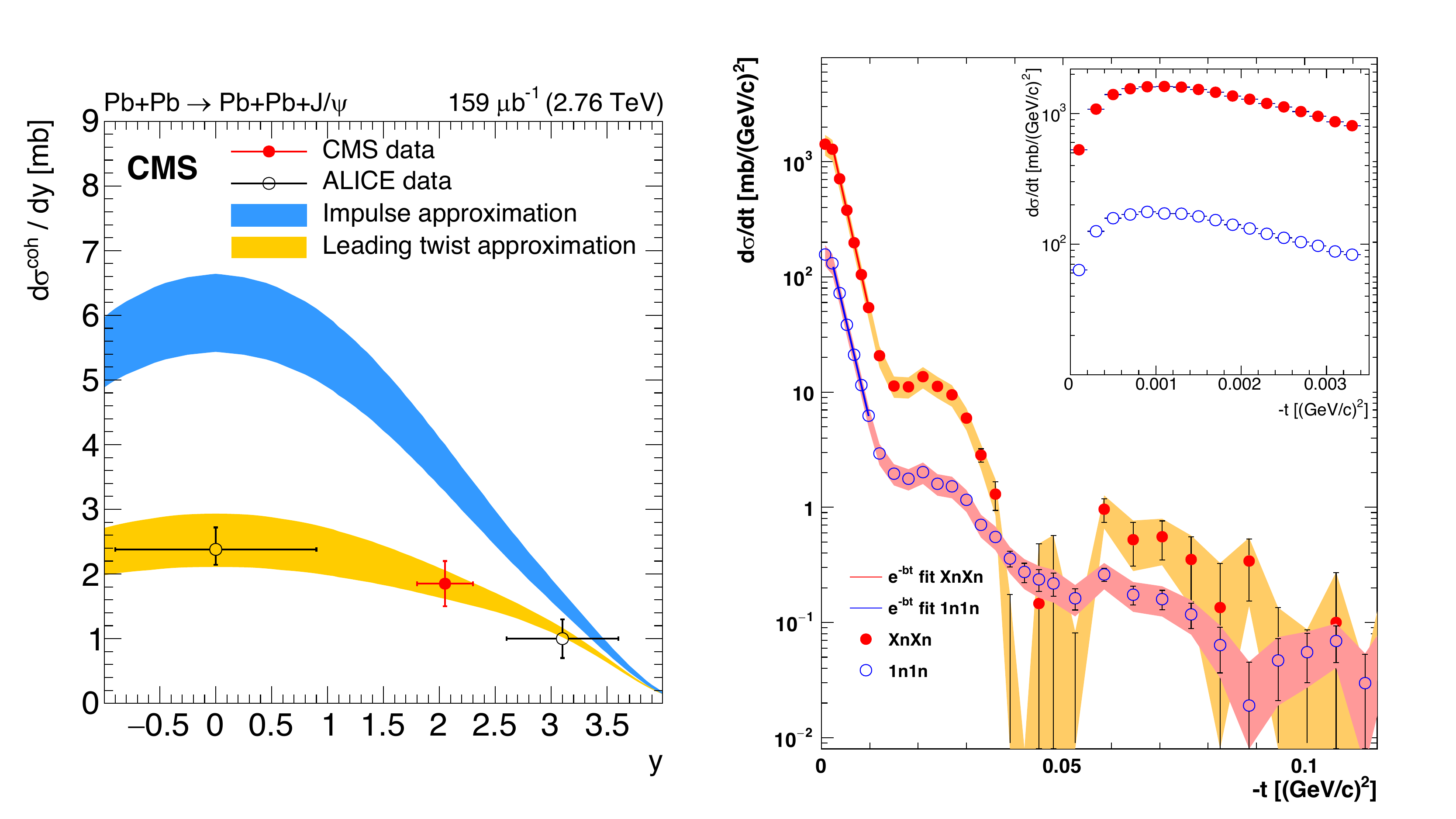}
\caption{(a) CMS and ALICE measurements of the \Jpsi\  $d\sigma/dy$ as a function of rapidity, for LHC Run 1 data. \cite{Khachatryan:2016qhq}.   (b) $d\sigma/dt$ for $\rho^0$ photoproduction in gold-gold collisions at a center of mass energy of 200 GeV/nucleon as measured by the STAR Collaboration \cite{Adamczyk:2017vfu}.  These data are for $\rho$ photoproduction accompanied by mutual Coulomb excitation; the two curves are for different selections of numbers of neutrons in each ZDC.  The inset shows $d\sigma/dt$ at very low $t$, where interference between the two directions is visible.}
\label{fig:cms-alice-jpsi}
\end{figure}

With these caveats in mind, it is useful to compare the heavy-ion data with pQCD calculations.  Fig. \ref{fig:cms-alice-jpsi} (a)  shows CMS central and ALICE forward muon spectrometer data on $d\sigma/dy$ 
%SKZ ($y$ is the final state rapidity) 
for $J/\psi$ photoproduction on a lead target.  Both are  about a factor of two below the impulse approximation, which treats the lead nucleus as a collection of free nucleons. It is, however, consistent with a `leading twist' pQCD calculation \cite{Guzey:2013qza,Frankfurt:2011cs}.   This is essentially an extension of the Glauber calculation discussed above which accounts for the possibility of the incident quark and antiquark to interact multiple times while traversing the nucleus, including excited intermediate states.  It treats shadowing as due to multiple scattering, without requiring changes in the actual parton densities. It will be very interesting to see data on shadowing in $\Upsilon$ photoproduction, where the $Q^2$ is larger.  Unfortunately, the larger background from $\gamma\gamma\rightarrow l^+l^-$ will be a bigger problem than in pPb or pp collisions.

For lighter mesons, like the $\rho$ and $\omega$, the $Q^2$ is low enough so that perturbative QCD is not expected to be applicable.  One can use $\gamma p$ data to predict the $\gamma A$ cross-sections with a Glauber calculation \cite{Frankfurt:2002sv}, as described above. 
Experiments observe a dipion mass spectrum with three components: $\rho^0\rightarrow\pi^+\pi^-$, $\omega\rightarrow\pi^+\pi^-$ and direct $\pi^+\pi^-$ production.  These three channels are indistinguishable, so they all interfere with each other. The direct $\pi^+\pi^-$ is flat, independent of mass, but, through interference, enhances the $\pi^+\pi^-$ spectrum below the $\rho$ mass and depletes the spectrum above it.  The branching ratio for $\omega\rightarrow\pi^+\pi^-$ is only 2.2\%, but, through its interference with the $\rho$,  the $\omega$ produces a kink in the mass spectrum, near the $\omega$ mass \cite{Adamczyk:2017vfu,Sirunyan:2019nog}.  The relative amplitudes of the three channels seem consistent with HERA data.  
Surprisingly, data on $\rho$ photoproduction, from both STAR \cite{Abelev:2007nb} and ALICE \cite{Adam:2015gsa}, shows a $\rho$ cross-section that is larger than that predicted by Glauber calculations, Eq. \ref{eq:glauber}.  One possibility is that nuclear inelastic scattering (by intermediate higher mass photon fluctuations) increases in the cross-section  \cite{Frankfurt:2015cwa}.

Photoproduction can also be used for vector meson spectroscopy.  Although meson photoproduction has long been studied in fixed target experiments \cite{Bauer:1977iq} and at HERA, current UPC analyses have collected large data samples (up to $\approx 1$M events) with high-quality detectors,  comparable in size and with larger maximum mass reach, and are starting to produce interesting results.  STAR has observed a $\pi^+\pi^-$ resonance with a mass around 1.65 GeV and a width around 165 MeV~\cite{Klein:2016mki}  The rate appears roughly consistent with photoproduction of the $\rho_3 (1690)$; this would be an interesting observation of a spin-3 meson at a $W_{\gamma p}$ where production via Pomeron exchange dominates. STAR \cite{Abelev:2009aa} and ALICE~\cite{Mayer2014} have also studied photoproduction of $\pi^+\pi^-\pi^+\pi^-$ final states, observing a broad resonance which seems consistent with a mixture of the $\rho(1450)$ and $\rho(1700)$ states. Data already collected, but not-yet analyzed, could be used to significantly improve our knowledge of heavier vector meson states. 

\subsection{The dipole model and nuclear imaging with coherent photoproduction}

An alternate approach to vector meson photoproduction treats the interacting photon as a quark-antiquark dipole with separation $r_T$.  This dipole may scatter in the target, emerging as a vector meson.  This approach treats protons and ions in a similar manner, via a target configuration $\Omega$ that describes positions of the gluons in the target.   It meshes smoothly with the Good-Walker approach to diffraction~\cite{Good:1960ba}, allowing calculations of incoherent photoproduction.  The cross-section to produce a vector meson is \cite{Kowalski:2006hc,Rezaeian:2012ji,Klein:2019qfb}
\begin{equation}
\frac{d\sigma}{dt} = \frac{1}{16\pi}\big|A^{\gamma A\rightarrow VA}\big|^2 (1+\beta^2),
\label{eq:dipolesigma}
\end{equation}
where 
\begin{equation}
A^{\gamma A \rightarrow VA} = i \int d^2 \vec{r}_T 
\int \frac{dz}{4\pi} \int d^2\vec{b}_T 
\psi_\gamma^*(\vec{r}_T,z,Q^2)
N_\Omega(\vec{r}_T,\vec{b}_T)
\psi_V(\vec{r}_T,z,Q^2)
e^{-i\vec{b}_T\cdot\vec{k_T}/\hbar},
\label{eq:dipoleamp}
\end{equation}
where $\vec{b}_T$ is the transverse position within the nucleus, $\vec{r}_T$ is the transverse size of the dipole, $z$ the fraction of the photon momentum carried by the quark (the antiquark has momentum fraction $1-z$) and $Q^2$ the photon virtuality.   The wave function of the incident photon, $\psi_\gamma^*(\vec{r}_T,z,Q^2)$ includes the probability of the photon fluctuating to a dipole. This is related to $\Gamma_{ll}$.  The wave function includes the probability distribution for the quark to carry a momentum fraction $z$.
%SKZ of the photon; 
This probability is symmetric around $z=0.5$, and weighted toward low mass dipoles, where $z\approx 0.5$. 

Different models have been used for the wave function of the outgoing vector meson, $\psi_V(\vec{r}_T,z,Q^2)$ \cite{Lappi:2013am}.  Quark models indicate that it should be Gaussian in $\vec{r}_T$.  The Gaussian-light-cone form is
\begin{equation}
\psi_V(\vec{r}_T,z,Q^2) = N [z(1-z)]^2 e^{-R_T^2/\sigma^2},
\end{equation}
where the normalization $N$ and width $\sigma$ come from are based on fits to data.  The ``Boosted Gaussian'' is slightly more complex, based on the Fourier transform of the (momentum-space) light cone wave function.

Here, $N_\Omega({\vec{r}_T,\vec{b}_T})$ is the imaginary part of the forward dipole-target scattering amplitude for a dipole with transverse size $\vec{r}_T,$ impacting the target at transverse position $\vec{b}_T$.   The small real part of the amplitude is accounted for by the $(1+\beta^2)$
term in Eq. \ref{eq:dipolesigma}.  The subscript $\Omega$ denotes the target configuration (nucleon positions, etc.). 
This formulation assumes that the dipole size is smaller than the nuclear target; if the dipole is larger the exponential becomes slightly more complicated. 

The optical theorem provides a simple relationship with the dipole scattering cross-section: 
$N_\Omega({\vec{r}_T,\vec{b}_T}) = d^2\sigma_{q\overline q}/d^2b$.  In a pQCD context, the cross-section depends on the gluon density $g(x,\mu^2)$ \cite{Mantysaari:2016jaz}
\begin{equation}
\frac{d^2\sigma_{q\overline q}}{d^2b} = 2\bigg(1-\exp{[-\frac{\pi^2}{2N_c}r^2 \alpha_s(\mu^2)xg(x,\mu^2T(b))]}\bigg).
\label{eq:dipolescat}
\end{equation}
Here, $T(b)$ is the thickness function - the integrated material encountered by a photon arriving at impact parameter $b$.   The exponential accounts for the probability of the dipole undergoing multiple interactions. For small dipoles, $d^2\sigma_{q\overline q}/d^2b \propto r_T^2$.

This dipole formulation has some limitations.  It assumes that the dipole does not change as it traverses the target. The lifetime of the fluctuation, $\hbar/M_{q\overline q}$ must be longer than the time spent in the nucleus, {\it i. e.} $E>\hbar c/R_A$, the same condition as in the pQCD formulation, but it is only implicit here.  It is usually satisfied at RHIC and the LHC, but can fail for high-mass final states, particularly at large rapidity for the lower photon-energy choice in Eq. \ref{eq:photonenergy}.  
  
Since the quark and antiquark momentum fractions are $z$ and $1-z$ respectively, there is no room for soft gluons; this is necessarily a lowest order formulation for the photon.  However, there is much more freedom in characterizing the target.  It is easy to do calculations with colored glass condensates or other saturation models by altering Eq. \ref{eq:dipolescat}.  The presence of significant shadowing leads to a narrowing of the $k_T$ distribution in the interaction, shifting the meson $\pT$ to smaller values \cite{Guzey:2016qwo}.

The dipole approach can also accommodate impact-parameter dependent variations in the cross-section. It has been used for a large number of different vector meson photoproduction calculations, using different wave functions and dipole-target cross-sections.   The cross-sections often use different models of gluon shadowing and/or saturation, including a different impact parameter dependence.  One expects more gluon shadowing in the core of the nucleus (small $\vec{b}_T$) than in its periphery \cite{Emelyanov:1999pkc}.  

Figure \ref{fig:alice-jpsi} (b) compares ALICE $J/\psi$ $d\sigma/dy$ lead-target data with predictions from a few representative calculations. 
\cite{Acharya:2019vlb}.  The impulse approximation treats the nucleus as a collection of independent nucleons, while $\SL$ uses a Glauber calculation based on parametrized HERA data.  BKG-I is a Glauber-Gribov calculation, which uses a gluon density extracted from HERA data \cite{Luszczak:2019vdc}.  Also shown is a leading-twist calculation (`LTA'), while the green line and shaded band show a pQCD calculation where gluons shadowing following the EPS09 nuclear shadowing parameterization, which is based on non-UPC data.   The figure displays three other dipole calculations with different models of the nucleus.  The 'IIM-BG' curve is based on a colored-glass condensate (CGC) model.  CGCs are a form of saturation model, whereby the nuclei are represented by a disordered classical gluon field. In the 'IPSat' (impact-parameter dependent saturation) model the dipole-proton cross-section depends on the dipole-proton impact parameter; the proton is represented with a Gaussian transverse matter distribution.  Finally, the GG-HS model includes gluon `hot spots,' as will be discussed below.  Aside from the impulse approximation, which, not surprisingly is far above the data, and the IIM BG calculation, which is considerably below it, all of the models are in at least marginal agreement with the data. The hot-spot and EPS09 calculations are are in broad agreement with the data at large rapidity (corresponding mostly to lower photon energies), but diverge for more central rapidities.  Fig. \ref{fig:cms-alice-jpsi}(a) shows this data, but the only theory comparison is with the LTA, where the agreement is good.  This is slightly surprising, since Fig. \ref{fig:alice-jpsi}(b) shows that the LTA curve tends to diverge from the data as $|y|$ is reduced.  There are many of other calculations of this process, with different treatments of gluon density in lead

Overall, most of these models do a reasonable job of matching the data.  Looking ahead, it will be very desirable to have a more complete set of comparison data, including the $J/\psi$, $\psi'$ and $\Upsilon$, all covering a wide range of rapidities, to more broadly test these models. 

\subsubsection{Nuclear Imaging}

Equation \ref{eq:dipoleamp}, $F(b)$ is the two-dimensional distribution of interaction sites in the target.  The same equation provides a way to fairly directly probe for changes in the nuclear profile, $F(b)$ due to shadowing. .  For protons, this offers a way to probe the generalized parton distributions \cite{Burkardt:2002hr}.  For UPCs, we focus on heavier ions.  The transformation is straightforward \cite{Toll:2012mb}:
\begin{equation}
F(b) \propto \int_0^\infty \pT d\pT J_0(b\pT)\sqrt{\frac{d\sigma}{dt}},
\label{eq:transform}
\end{equation}
where $J_0$ is a modified Bessel function. 

%SKZ following sentence modified since it didn't make sense. 
There are significant difficulties when this is put into practice, particularly for ions. The relationship is exact if the integral is unbounded, and $bp_T$ must cover several cycles of the Bessel function for an accurate transform. However, the coherent cross-section drops faster with increasing $t$ than both incoherent production and the background, so the data has a limited maximum useful $\pT$.  A cutoff at finite $\pT$ introduces windowing artifacts in the transform; the data is effectively convolved with a box function \cite{Adamczyk:2017vfu}, so the calculated $F(b)$ includes the box function as well.  The equation also assumes that $p_T$ is the Pomeron transverse momentum, but the measured $p_T$ also includes the photon $p_T$. 
For ion targets, $d\sigma/dt$ shows diffractive minima, e.g. as shown by STAR data~\cite{Adamczyk:2017vfu} in Fig. \ref{fig:cms-alice-jpsi} (b), which signals a sign change in the photoproduction amplitude.  In Eq. \ref{eq:transform}, $\sqrt{d\sigma/dt}$ is this amplitude, so it is necessary to flip its sign when crossing each minimum.  In UPCs, these minima are smeared out because of the photon $\pT$, so determining the dip positions is not straightforward.    Studies of the dipion mass (a proxy for $Q^2$) dependence of $F(b)$ in dipion photoproduction show an intriguing trend, but the systematic uncertainties are large \cite{Klein:2018grn}.  Other approaches for measuring the change in shape of $d\sigma/dt$ may be more promising. 

\subsection{Incoherent Photoproduction}

\begin{figure}[t]
\includegraphics[width=0.97\textwidth]{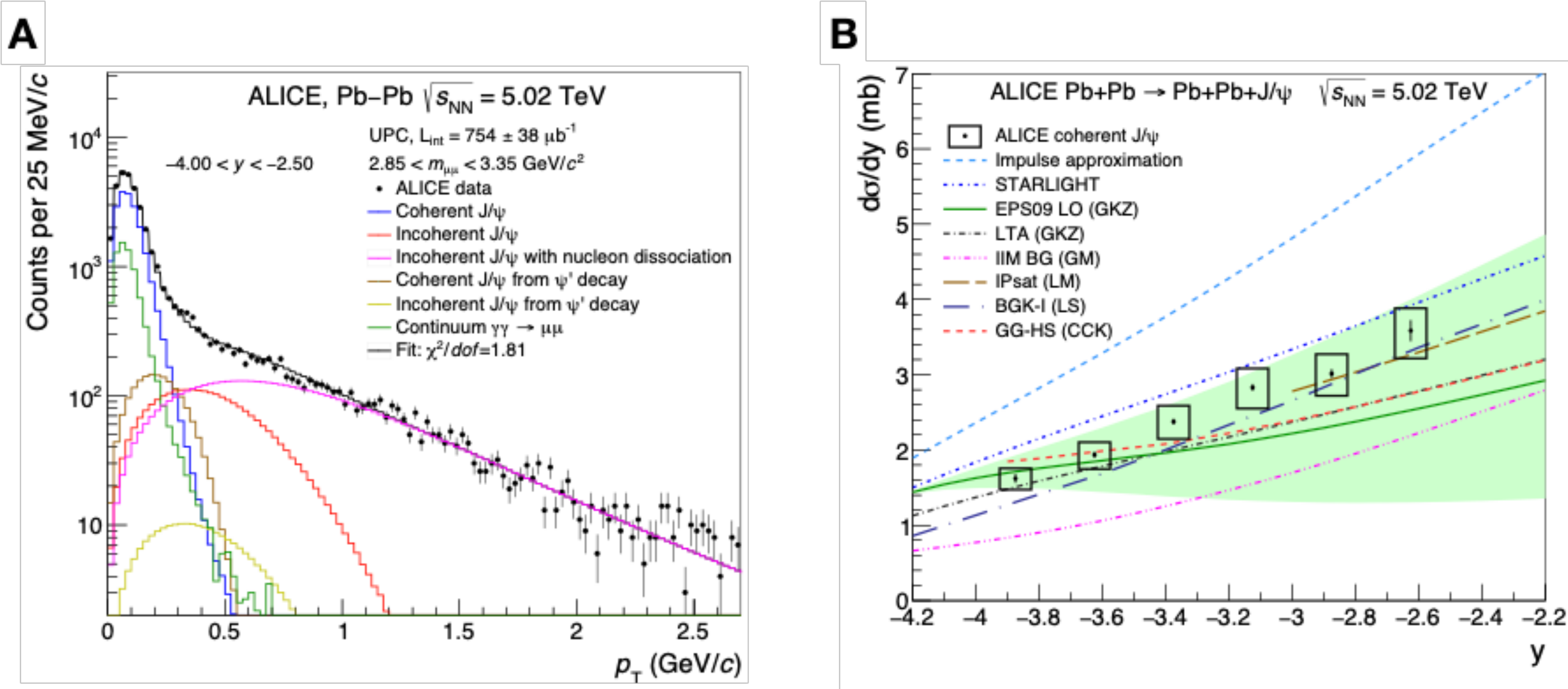}
\caption{(a) Dimuon $\pT$ distribution for dielectron pairs in the $\Jpsi$ mass window, (b) Differential cross section $d\sigma/dy$ for UPC events in 5.02 TeV Pb+Pb data. From Ref. \cite{Acharya:2019vlb}.}
\label{fig:alice-jpsi}
\end{figure}

While coherent photoproduction probes the average nuclear configuration, incoherent photoproduction is sensitive to fluctuations, both spatial nuclear fluctuations and local fluctuations in the gluon density.  This relationship comes from the optical theorem, and is usually embodied in the Good-Walker 
formalism for diffraction \cite{Good:1960ba}. In the Good-Walker approach, the total cross-section is \cite{Klein:2019qfb}
\begin{equation}
\frac{d\sigma_{\rm tot}}{dt} = \frac{1}{16\pi} \langle|A_\Omega|^2\rangle,
\end{equation}
where the added $\Omega$ subscript on $A$ from Eq. \ref{eq:dipoleamp} explicitly shows that it depends on the target nuclear configuration $\Omega$.
%SKA it's not just the spatial configuration here
In coherent scattering, the initial and final nuclear states are the same, so the amplitudes for the different states are summed and then squared, and
\begin{equation}
\frac{d\sigma_{\rm coherent}}{dt} = \frac{1}{16\pi} |\langle A_\Omega\rangle|^2.
\end{equation}
Because the averaging occurs before squaring, coherent photoproduction is sensitive to the average of the configurations. The incoherent cross-section is the remainder after the coherent cross-section is subtracted from the total.  
\begin{equation}
\frac{d\sigma_{\rm incoherent}}{dt} = \frac{1}{16\pi} 
\big(\langle|A_\Omega|^2\rangle -|\langle A_\Omega\rangle|^2\big).
\end{equation}
Through the first term, the incoherent cross-section is sensitive to event-by-event fluctuations in the target configuration \cite{ Miettinen:1978jb}.  It requires that the target have an internal structure; without internal structure, there are no inelastic interactions. 

One consequence of the Good-Walker approach is that, at very high energies, incoherent photoproduction should disappear.  The photonuclear cross-section rises with photon energy, as 
photoproduction occurs on gluons with smaller and smaller $x$ values. When the photonuclear cross-section is large enough, the nucleus looks like a black disk.  At that point, the internal structure disappears, and the incoherent cross-section vanishes~\cite{Frankfurt:2005mc}.

Theorists have used incoherent HERA data on $J/\psi$ photoproduction to study proton shape fluctuations  \cite{Mantysaari:2016jaz}. Their analysis found that the cross-section for incoherent photoproduction was above the expectations for a smooth proton, but was consistent a model where the proton contained  regions of high gluon density (`hot spots').   The number of hot spots should increase with photon energy, as the target gradually becomes opaque.   In one calculation, the incoherent $J/\psi$ photoproduction cross-section increases with energy, reaching a maximum at $W_{\gamma p}=500$ GeV, and then decreases with further increases in $W_{\gamma p}$ \cite{Cepila:2016uku}, within the reach of LHC data. The same calculation found that the energy at which the incoherent maximum appears increases with vector meson mass, so it is only about $W_{\gamma p}=20$ GeV for the $\rho$.

Similar approaches can be applied to nuclei \cite{Cepila:2017nef}. The hot spots are the same as in protons and the number of hot spots in a nucleus is $A$ times larger than in a proton.  Although the differences in inelastic cross-sections between the two models are smaller than for proton targets, the hot spot model predicts a larger inelastic $J/\psi$ production, with the difference rising with increasing $t$.  Because saturation sets in at much lower $W_{\gamma p}$ for the $\rho$, the incoherent $\rho$ photoproduction cross-section is predicted to be a small fraction (less than 5\%) of the coherent $\rho$ cross-section.  That prediction is in some tension with STAR data, where the ratios inferred from integrating $d\sigma/dt = A \exp(-bt)$ from Refs.  \cite{Abelev:2007nb} and Ref. \cite{Agakishiev:2011me} seem to be considerably higher.

Experimentally, the distinction between coherent and incoherent is not completely straightforward for nuclei.   The STAR data in Fig. \ref{fig:cms-alice-jpsi}(b) is actually from the reaction $AA\rightarrow A^*A^*\rho^0$; the STAR trigger required neutron emission from each nucleus.  In principle, the Good-Walker requirement for coherent photoproduction, that the initial and final states are the same, is not satisfied.  However, the data shows a clear coherence peak for $\pT<\hbar/R_A$, and at least one diffractive minimum.  This may be explained using Eq. \ref{eq:dipoleamp}, where coherence depends only on the transverse position of the nucleons.  Nuclear excitation is a relatively soft process, and the time scales for it to affect the target nucleus are much longer than the time scales required to produce the vector meson.  If the excitation is caused by an additional photon, then it does not destroy the coherence of the vector meson photoproduction.  The process essentially factorizes \cite{Baur:2003ar,Baltz:2002pp}.  The STAR and ALICE $\rho$ cross-section measurements show that this factorization seems to hold quite well \cite{Adler:2002sc,Abelev:2007nb,Acharya:2019vlb}, except possibly at large $\pT$.   

If the presence or absence of neutrons does not completely determine whether a reaction is coherent or incoherent, the coherent cross-section may be found by fitting the incoherent component at large $\pT$, extrapolating down to small $\pT$ and subtracting.  The accuracy of this method depends on the functional form that is used. Although an exponential function has frequently been used to model the incoherent $d\sigma/dt$, a high-statistics analysis showed that an exponential is not a good fit to the data.  A dipole form factor provides a better match \cite{Klein:2018grn}.  Alternately, the form can be derived from Monte Carlo simulations \cite{Acharya:2019vlb} which include the photon $p_T$.  

A recent ALICE study \cite{Acharya:2019vlb} went further, and divided the incoherent interactions into two classes.  In the first, the nucleus dissociated, but the individual nucleons remained intact.  In the second, the individual nucleons were excited into higher states.  For the first class, they used a $\pT$ template from the \SL\ Monte Carlo \cite{Klein:2016yzr}.  For the second, they used a parametrization obtained by HERA studies of the same type of reaction.   As Fig. \ref{fig:alice-jpsi}(a) shows, this combination provides a very good fit to the measured $J/\psi$ $p_T$ spectrum. 

Even without a trigger that requires neutrons, separating coherent and incoherent interactions is not simple, especially for heavy ions.  Since $Z\alpha\approx 0.6$, additional photons may be exchanged, exciting one of the nuclei. It is impossible to tell whether one photon incoherently produced a vector meson, or one photon coherently produced that meson, and a second photon dissociated the nucleus.  

\subsubsection{Prospects for measuring GPDs with collisions involving polarized protons}

Because it can accelerate polarized protons, RHIC can uniquely study polarized generalized parton distributions (GPDs), which describe where the partons are within a nucleon (i. e. as a function of $b$, similar to Eq. \ref{eq:transform}, but for protons). Polarized GPDs are sensitive to  parton polarization~\cite{Accardi:2012qut}.   The primary experimental observable is the single spin asymmetry~\cite{Lansberg:2018fsy}, which is proportional to the GPD $E^g$, which quantifies the gluon orbital momentum.
The STAR Collaboration performed an initial measurement of this observable, albeit with a large statistical uncertainty \cite{Schmidke2019}.  They used $pA$ collisions, where photons from the gold nucleus illuminated the polarized proton target.  The proposed AFTER experiment plans to study this process with unpolarized lead ions striking a polarized proton target \cite{Lansberg:2018fsy}. 

\subsection{Photoproduction of exotic hadrons}

Reggeon exchange allows a wide range of final states, since spin and charge can be exchanged   Because the spin or charge exchange alters the target, coherence is unlikely to be maintained, so most studies of Reggeon exchange have considered proton targets.  Reggeon exchange rates are high.  The predicted photoproduction rate for the $a_2^+(1320)$, a `standard candle' $q\overline q$ meson is about a billion mesons/year, at both RHIC and the LHC \cite{Klein:2019avl}.  So, states with small couplings to photons can be observed. The cross-sections peak at 
low (a few times threshold) photon energies.  These low photon energies correspond to large-rapidity final states.  A given final state will be nearer mid-rapidity at RHIC than at the LHC.  Measurements at RHIC will provide good opportunities \cite{Klein:2019avl}, especially with future forward upgrades such as the STAR Forward Tracking and Calorimeter Systems  \cite{Yang:2019bjr}.

UPC photoproduction is a way to study the exotic $XYZ$ states \cite{Brambilla:2019esw}.  These states are heavy, containing a $c\overline c$ pair, so are mostly beyond the reach of fixed-target photoproduction. Photoproduction data would help elucidate the nature of these states \cite{Lin:2013ppa,Wang:2015lwa,Goncalves:2014iea}.   Pentaquarks may also be produced, via $\gamma p\rightarrow P$ \cite{Goncalves:2019vvo}.  Jefferson Lab's GlueX experiment has recently put strong limits on  photoproduction of $c\overline c$ containing pentaquarks \cite{Ali:2019lzf}. UPCs should be useful to study heavier pentaquarks, such as those containing $b\overline b$ pairs.

\subsubsection{Photoproduction in Peripheral Hadronic Collisions}

Although the criterion $b>2R_A$ impacts its visibility (and provides the largest rate, via $Z^4$),  photoproduction still occurs when $b<2R_A$.  Both STAR \cite{STAR:2019yox} and ALICE\cite{Adam:2015gba} have observed an excess, over hadronic expectations, of $J/\psi$ with $p_T<150$ MeV/c in peripheral collisions. The cross-section for these $J/\psi$ generally agrees with photoproduction calculations \cite{Klusek-Gawenda:2015hja,Zha:2017jch,GayDucati:2018who}. These calculations raise an interesting question: Do nucleons that interact hadronically also contribute to the photoproduction amplitude? If they lose energy, the photoproduction cross-section will be drastically reduced.  This affects both the $p_T$ distribution of the $J/\psi$ as well as their abundance.  The $J/\psi$ survival probability also merits study.  A $J/\psi$ produced inside the hadronic fireball may be dissociated before it can decay.  Even if it is produced outside the fireball, with its low transverse velocity and long lifetime (compared to the fireball), it may be engulfed before it decays.  High-statistics studies of how the cross-section depends on $b$ could shed light on these questions.

Photoproduced $J/\psi$ can also aid our understanding of the hadronic collision, by providing an independent (from the hadronic interaction) measurement of the reaction plane.  
%SKZ - some rewriting here to make it sound better.
Two variables have some sensitivity to the reaction plane. 
First, Eq. \ref{eq:interfere}, which relates the $J/\psi$ $p_T$ spectrum to the
angle between the $J/\psi$ $p_T$ and to $\vec{b}$.  The $J/\psi$ polarization provides additional information, since the $J/\psi$ linear polarization follows the photon polarization, which follows its $\vec{E}$ field, which is correlated with $\vec{b}$.  This polarization may be observed by azimuthal angle of the decay lepton $p_T$. Their transverse momenta tend to follow the $J/\psi$ polarization \cite{Baur:2003ar}. These handles depend on the direction of $\vec{b}$, and, from that, the reaction plane.   

\section{Two photon interactions}

The large photon fluxes from each nucleus, which each scale as $Z^2$, provide a high rate of photon-photon collisions, spanning a wide kinematic range, particularly at the LHC where the Lorentz factor is quite large.  Photon interactions lead to a wide variety of final states. They couple to all charged particles, including leptons, quarks, and charged gauge bosons~\cite{Baur:2001jj,Hencken:1995me}.  They also couple to neutral final states through loop diagrams, and so offer the possibility of observing direct production of Higgs bosons as well as two-photon final states (now also known as `light by light' scattering~\cite{dEnterria:2013zqi}).

\subsection{Two-photon luminosity} 

The production rates for exclusive particle production from two photon collisions can be factorized into a two photon "luminosity" (or flux) and the cross section ($\sigma(\gamongam \rightarrow X$)~\cite{Brodsky:1971ud,Budnev:1974de}.  The ultraperipheral two photon flux (often referred to as the ``two photon luminosity") is typically calculated by integrating over the two separate fluxes, with the requirements that 1) the nuclei do not overlap, using $\Pnohad$, 2) that the production does not take place inside either nucleus, 3) requiring a specific forward neutron topology selection (or ignoring this for an inclusive selection)~\cite{Baltz:2009jk,Klein:2016yzr}:

\begin{equation}
    \frac{d^2 N}{dk_1 dk_2} =
    \int_{b_1 > R_A} d^2 \vec{b}_1 
    \int_{b_2 > R_A} d^2 \vec{b}_2 
    N(k_1, b_1) N(k_2, b_2)
        \Pnohad ( |\vec{b}_1 - \vec{b}_2| )
        \Pzdc ( |\vec{b}_1 - \vec{b}_2| ).
\end{equation}
This integral can be simplified by changing variables, to integrate over the absolute values of $b_1$ and $b_2$ and the angle between them \cite{Cahn:1990jk,Baur:1991fn}. This approach is used in most calculations.  The second assumption, that there is no production inside either nucleus, 
is perhaps too strong, although the contributions to the flux should be limited due to the rapid fall off of the field strengths inside a nucleus. 

\subsection{Dilepton production: $\gamongam \rightarrow \ell+\ell$}

Most existing measurements of \gamongam\ processes in heavy ion collisions involve leptonic final states $\gamongam \rightarrow \ell^+\ell^-$.  At lower pair masses, $M<10$ GeV, they are typically performed in tandem with the vector meson measurements described previously~\cite{Afanasiev:2009hy,Abbas:2013oua}. 
The exclusive final state is particularly simple, consisting primarily of two back-to-back charged particles with opposite sign, and many high-energy and nuclear physics experiments have been designed with capabilities including excellent lepton identification and precise momentum measurements.  STAR has performed a series of electron pair measurements using their large acceptance time projection chamber (TPC), initially triggered by a mutual Coulomb exchange (neutrons in  both ZDCs, see Section 2.2), and confirmed by signals from the TPC and time-of-flight counters, to select events with electron pair candidates~\cite{Adam:2018tdm,Adams:2004rz}. 
ATLAS utilizes a sophisticated multi-level trigger system to select events with a single muon, vetoing on large transverse energies in the rest of the event, but with no selection on the forward neutron topology~\cite{ATLAS:2016vdy}.  

\begin{figure}[t]
\includegraphics[width=0.97\textwidth]{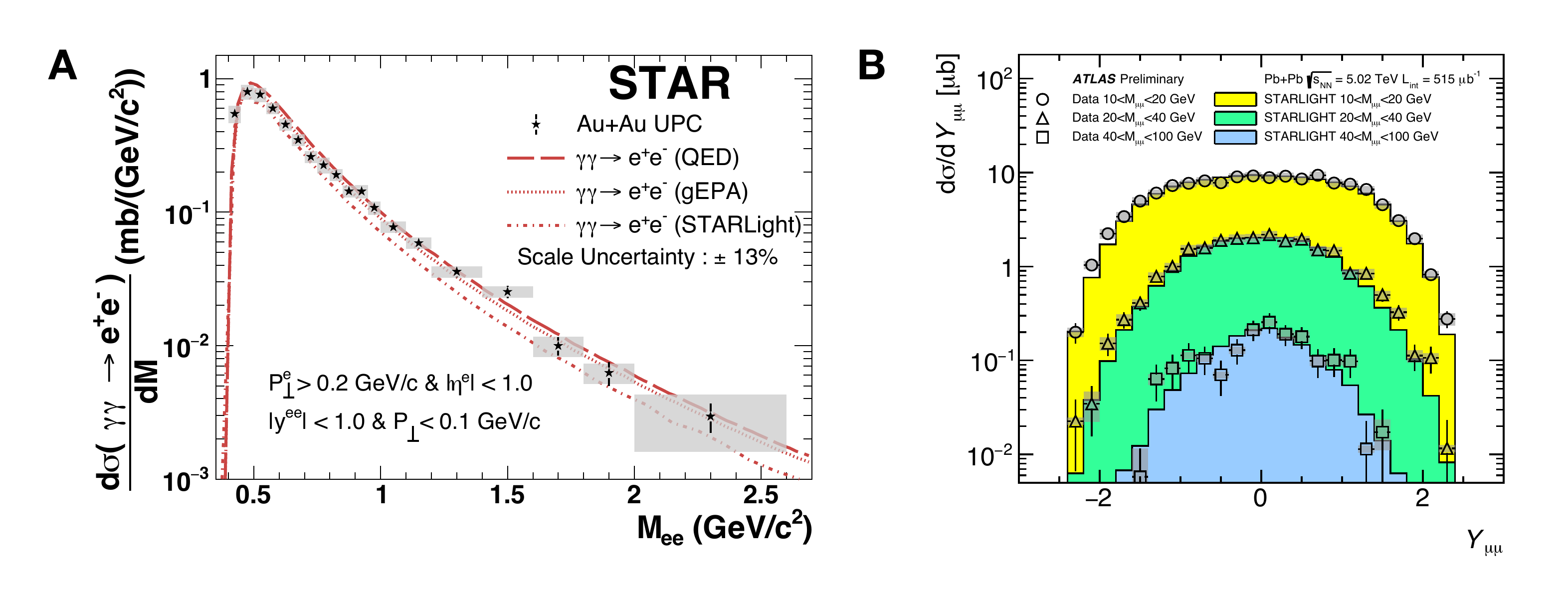}
\caption{(a.) Distribution of dielectron invariant mass from STAR~\cite{Adam:2019mby}. (b.) Distribution of dimuon rapidity, in invariant mass selections, from ATLAS~\cite{ATLAS:2016vdy}.}
\label{fig:star-atlas-ll-xsec}
\end{figure}

Dilepton cross sections are shown in Fig.~\ref{fig:star-atlas-ll-xsec} from STAR~\cite{Adam:2019mby} (as a function of invariant mass) and ATLAS~\cite{ATLAS:2016vdy} (as a function of pair rapidity, for three mass ranges), and compared to calculations including $\SL$~\cite{Klein:2016yzr}.  ATLAS restricts the measurements to relatively large muon transverse momenta ($p_{T\mu} > 4$ GeV in $|\eta| < 2.4$), which also selects large mass (ATLAS chose $M_{\mu\mu}>10$ GeV), but has no restriction on the forward neutron topology, and thus measures the full fiducial cross section.  STAR measures much softer electrons ($p_{Te}>0.2$ GeV in $\eta < 1$) and thus much lower masses, but its UPC pairs are required to have forward neutrons in both ZDCs, which limits the cross section to only a small fraction of the total.  \SL\ describes the overall magnitude of the dilepton cross section over a wide range in pair mass and rapidity, although ATLAS sees an underprediction from it in the forward region, and STAR sees perhaps some overall underprediction of the data.  

Although generalized EPA and full QED calculations~\cite{Zha:2018tlq} describe the magnitude of the cross-section better than $\SL$, STAR reports that all calculations have been found to be consistent with the data within the stated overall scale uncertainties of 13\%.  The STAR data is for $e^+e^-$;  here the Weizsacker-Williams approach may be more questionable than for heavier leptons, because, uniquely, $m_e < \hbar/R_A$. Other groups~\cite{KlusekGawenda:2010jq,Sengul:2015ira,Azevedo:2019fyz,Vysotsky:2018slo} have performed calculations of the exclusive dilepton cross sections, albeit with somewhat different assumptions on the impact parameter dependence of the absorptive corrections, or by explicit inclusion of the nuclear form factors. 
The details of these features can have observable consequences on the magnitudes of the cross section.  In particular, the detailed shape of the pair spectrum at large values of either the pair invariant mass and pair rapidity distributions are both sensitive to the higher energy photons in the initial state.  Allowing pair production within the two nuclei improves agreement with the STAR data \cite{Zha:2018ywo}, and could improve the agreement with the ATLAS data.

STAR has also observed angular modulations of the pair momentum relative to the single electron momenta, reflecting the linear polarization of the initial photons~\cite{Li:2019yzy,Adam:2019mby}.

Tau ($\tau$) leptons can be pair-produced only for incoming diphoton invariant masses above $3.5$ GeV, so they are difficult to produce at RHIC.  However, they have been observed to exist at the LHC through leptonic decays where one decays to an electron and the other to a muon~\cite{ATLAS-EVENTDISPLAY-2018-009}, and should also be observable as low activity events containing one lepton and one or three charged tracks.  The coupling between the $\tau$ and photon is sensitive to a combination of modifications of $a_\tau = (g_\tau - 2)/2$, e.g. due to lepton compositeness~\cite{Silverman:1982ft} or coupling to supersymmetric particles~\cite{Martin:2001st}, as well as an electric dipole moment (EDM) of the $\tau$ itself.  This is predicted to have an observable effect on the $\pT$ of the decay leptons and hadrons, with a systematic hardening of the spectrum correlating with changes in $a_\tau$ or the $\tau$ electric dipole moment $d_\tau$~\cite{Beresford:2019gww}.  

\subsection{Dilepton \pT\ and impact parameter selections}

As has been emphasized throughout this 
%SKZ report
review, and in particular Sec.~\ref{subsec:pt}, one 
key feature which distinguishes production processes in purely electromagnetic processes from heavy ion collisions is the very low $\pT$ of the initial state photons.  In dilepton processes, this leads to final states pair $\pT \sim  20-30$ MeV. The pair $\pT$ can be measured accurately for low $\pT$ leptons (e.g. in STAR) but only estimated through the dilepton opening angle for high $\pT$ leptons (e.g. the acoplanarity, $\alpha = 1 - |\Delta\phi|/\pi$), as measured in ATLAS.  The detailed shapes of these distributions are sensitive to several different aspects of the underlying physics.

Large values of pair $\pT$ or $\alpha$ are generally inaccessible in most calculations, which typically provide the intrinsic $\pT$ 
via the nuclear form factor.  This part of the dilepton angular spectrum is generally sensitive to final-state higher-order photon emissions, as shown diagrammatically in Fig.~\ref{fig:Feynman} (f). These have generally not been included in existing event generators, although Refs.~\cite{Klein:2018fmp,Klein:2020jom} have demonstrated analytically, using a Sudakov formalism, that they are already consistent with the preliminary ATLAS data.  Similar success modeling the experimental data should be possible using the final-state QED parton-showering already available in \PY\ 8~\cite{Sjostrand:2014zea}.  \PY\ also incorporates 
nuclear photon fluxes~\cite{Helenius:2018mhx}, albeit without overlap removal.

By contrast, pairs with small pair $\pT$ or $\alpha$ have been found to be particularly sensitive to quantum interference effects related to constraints on the impact parameter between the two colliding nuclei.
Up to this point, all processes have been assumed to exclude the fraction of the cross section where the nuclei overlap as well as when the production vertex is inside either nucleus~\cite{Baltz:2009jk}. This choice is made both out of convenience and to alleviate potential conceptual issues. Distinguishing signal processes from backgrounds is generally more difficult in events where hadronic processes also occur. There are also open questions as to where a coherent process can take place, in proximity to more violent hadronic collisions.
A subtler issue arises  from the limits of the equivalent photon approximation.  When comparing typical EPA calculations with those using a more complete QED formalism, replacing the photon probability densities with QED amplitudes based on point charge nuclei convolved with measured form factors,
These integrals involve a phase $i\vec{b}\cdot\vec{q}$, where $\vec{q}$ is the vector pair transverse momentum \cite{Vidovic:1992ik,Vidovic:1993cf,Zha:2018tlq}. Restricting the impact parameter close to zero can lead to large oscillations that tend to deplete the cross sections for low pair $\pT$ and lead to increased broadening in the final $\pT$ or $\alpha$ distributions. This effect was first observed in dielectron data from STAR~\cite{Adams:2004rz}, by comparison to QED calculations from Hencken et al.~\cite{Hencken:2004td} which evinced a clear suppression at low pair $\pT$ relative to EPA calculations.

\begin{figure}[t]
\includegraphics[width=0.97\textwidth]{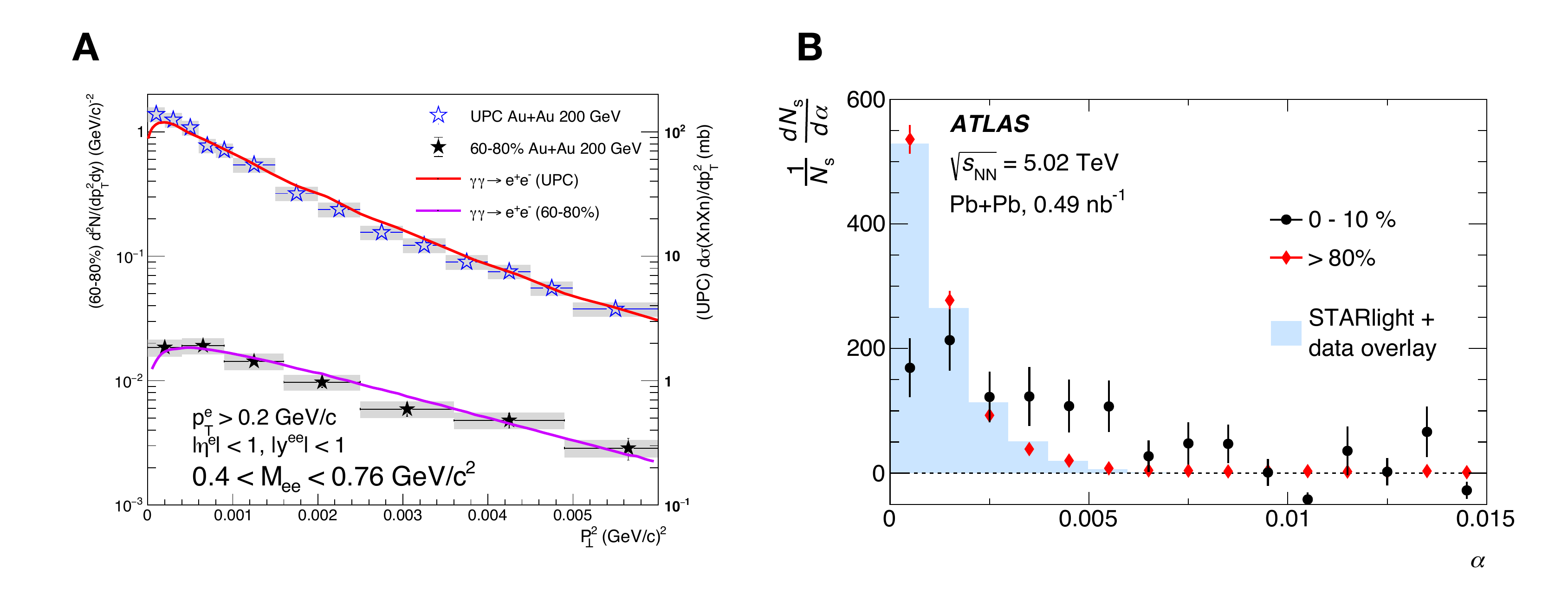}
\caption{(a.) STAR data on dielectron $\pT$ distributions in UPC and peripheral Au+Au collisions~\cite{Adam:2019mby}. (b.) ATLAS comparisons of dimuon acoplanarity ($\alpha$) between UPC and central collisions~\cite{Aaboud:2018eph}.}
\label{fig:star-atlas-nonupc}
\end{figure}

More recently, STAR studied data from Au+Au and U+U collisions measured with a UPC selection, and for very peripheral (60-80\%) hadronic interactions~\cite{Adam:2018tdm,Adam:2019mby}.  They observe a distinct broadening of the $\pT$ distributions, illustrated in Fig.~\ref{fig:star-atlas-nonupc} (a), an effect that had been postulated to be evidence for trapped magnetic fields but which was revised based on the existence of updated calculations in Ref.~\cite{Zha:2018tlq}.
Similarly, ATLAS observed nearly-back to back muon pairs in hadronic Pb+Pb collisions at the LHC, but over the full (0-100\%) centrality range.  They observed a clear centrality-dependent broadening of the $\alpha$ relative to a UPC-like selection~\cite{Aaboud:2018eph}, as shown in Fig.~\ref{fig:star-atlas-nonupc} (b).  This was interpreted by ATLAS and other authors as a possible probe of the charged constituents of the hot, dense quark gluon plasma.  However, a preliminary update of this measurement with more than three times the integrated luminosity ~\cite{ATLAS-CONF-2019-051} observed that the acoplanarity distributions were not just broadened in more central collisions, but peaked at non-zero values.  Intriguingly, the QED calculations in Ref.~\cite{Zha:2018tlq} seems to have predicted the dip at zero, and quantitatively described a large fraction of the full distribution, suggesting that no exotic QCD physics is needed, but rather a more careful treatment of QED interference effects, related to constraints on the impact parameter range.

\subsection{Bound-free pair production, antihydrogen production and accelerator luminosity limits}

Pair production does not always result in a free electron; the electron may be produced bound to one of the incident ions \cite{Bertulani:1987tz}.  This is known as bound-free pair production (BFPP).  The cross-section for BFPP with the electron captured into the $K$ shell is \cite{Aste:2007vs}
\begin{equation}
\sigma = \frac{33\lambda_{c}^7 Z_T^5 Z_p^2\alpha^7 }{20(e^{2\pi\alpha Z}-1)},
\big(\ln{\frac{\delta\gamma}{2}}-\frac{5}{3}\big)
\end{equation}
where $Z_p$ and $Z_T$ are the charge of the photon emitting nucleus and target nucleus respectively, $\lambda_c$ is the Compton wavelength and $\delta\approx0.681$.
The flux scaling with $Z_p^2$ is standard for photon emission, while $Z_T$ enters at the 5th power because the depth and width of the electric potential well increase rapidly with $Z$.  The total BFPP cross-section is about 20\% larger, because of the possibility for capture into higher orbitals.  The cross-section is maximal for relatively low-energy photons (2.5 MeV in the target rest frame), so the ion momentum is largely unchanged, despite the charge change.
% process has several interesting applications.

The cross-section for BFPP is large,  about 276 barns for each target beam for lead-lead collisions at the LHC \cite{Baltz:1996as}.  Because the produced single-electron ions have a larger magnetic rigidity, they are lost from the beam.  Along with Coulomb excitation of ions, these single electron ions are a major source of beam and luminosity loss over time. 

BFPP at the LHC produces a fairly well collimated beam of  single-electron lead ions which diverge from the circulating fully-stripped lead ions \cite{Klein:2000ba,Jowett:2016yfa}.  The trajectory of this beam depends on the the LHC magnet optics, but it will strike the beampipe more than 100 m downstream of each interaction point. This beam carries significant power which produces local heating that may cause the superconducting magnets to quench.  In a controlled test, a magnet quench  occurred at a luminosity of $2.3\times 10^{27}$ cm$^{-2}$ s$^{-1}$, 2.3 times the design luminosity \cite{Jowett:2016yfa}.  Although the local heating problem can be alleviated by spreading out the single-ion beam via orbit bumps and/or providing increased cooling, BFPP is an important constraint on high-energy accelerator design for heavy ions.  

BFPP can also produce antihydrogen: positrons bound to antiprotons.  This process was used to produce the first antihydrogen atoms at the CERN's Low Energy Antiproton Accumulator (LEAR) ring \cite{Baur:1995ck} and then at the Fermilab Antiproton Accumulator \cite{Blanford:1997up}.  Detection is easier with low energy collisions, where the antihydrogen velocity is relatively modest.  

Similar reactions occur with muons \cite{PhysRevA.53.1498}, taus and even charged mesons. The cross-section for muonic BFPP is has been calculated to be 0.16 mb  \cite{Bertulani:2010na}; the cross-section for heavier particles is smaller, less than 100$\ \mu$b.  Since these single-lepton atoms have large rapidities, they can only be detected with a far-forward spectrometer.

It is also possible to produce the bound state $\mu^+\mu^-$ \cite{Azevedo:2019hqp}.  They are produced near mid-rapidity. In lead-lead collisions at the LHC, the cross-section is $\approx 1\mu$b, so the production rate is large; the difficulty is in detecting the two soft photons in the para-muonium final state.

\subsection{Light-by-light scattering: $\gamongam \rightarrow \gamongam$}

Classically, QED obeys the principle of superposition, such that electromagnetic fields are purely additive, so photons do not interact with each other.   However, the standard model predicts~\cite{dEnterria:2013zqi,Klusek-Gawenda:2016euz} that photons can interact via loop diagrams, with internal lines containing quarks, leptons, and charged W gauge bosons, as shown in Fig.~\ref{fig:Feynman}. The cross-section is sensitive to beyond standard model processes such as magnetic monopoles~\cite{Ginzburg:1998vb}, vector fermions~\cite{Fichet:2014uka}, and axion-like particles (ALPs)~\cite{Knapen:2016moh,Bauer:2017ris}.  Related processes have been observed by Delbr{\"u}ck scattering off the coulomb field of a nucleus~\cite{Jarlskog:1974tx}, as well as photon splitting~\cite{Akhmadaliev:2001ik}, but the direct process was only recently observed, with greater than 5$\sigma$ significance, by the ATLAS experiment~\cite{Aad:2019ock} using 1.7 nb$^{-1}$ of Pb+Pb data at 5.02 TeV, after earlier evidence published by ATLAS~\cite{Aaboud:2017bwk} and CMS~\cite{Sirunyan:2018fhl} with only 0.4-0.5 nb$^{-1}$.

ATLAS~\cite{Aad:2019ock} utilized photons measured in their electromagnetic calorimeters, rejecting backgrounds by requiring no tracks in the tracking detectors matched to the photon. The primary fiducial selection is to have each photon with $E_\mathrm{T}>3$ GeV and a maximum $|\eta|<2.4$, approximating the acceptance of the electromagnetic calorimeter, and an invariant mass above 6 GeV.   Before the correction factor is applied, backgrounds are removed both by data driven techniques estimating the probability of mis-tagging electrons as photons, as well as by a data-driven normalization of diphoton events produced through central exclusive production in a two gluon exchange.  
After all selections, 59 events were found in the signal regions, where $12\pm3$ background events were expected.
The significance of this result, relative to a background only hypothesis, was evaluated for $A_\phi<0.005$,
with 42 signal-region events, and $6\pm2$ background events are expected, giving a significance of 8.2 
$\sigma$.

\begin{figure}[t]
\includegraphics[width=0.97\textwidth]{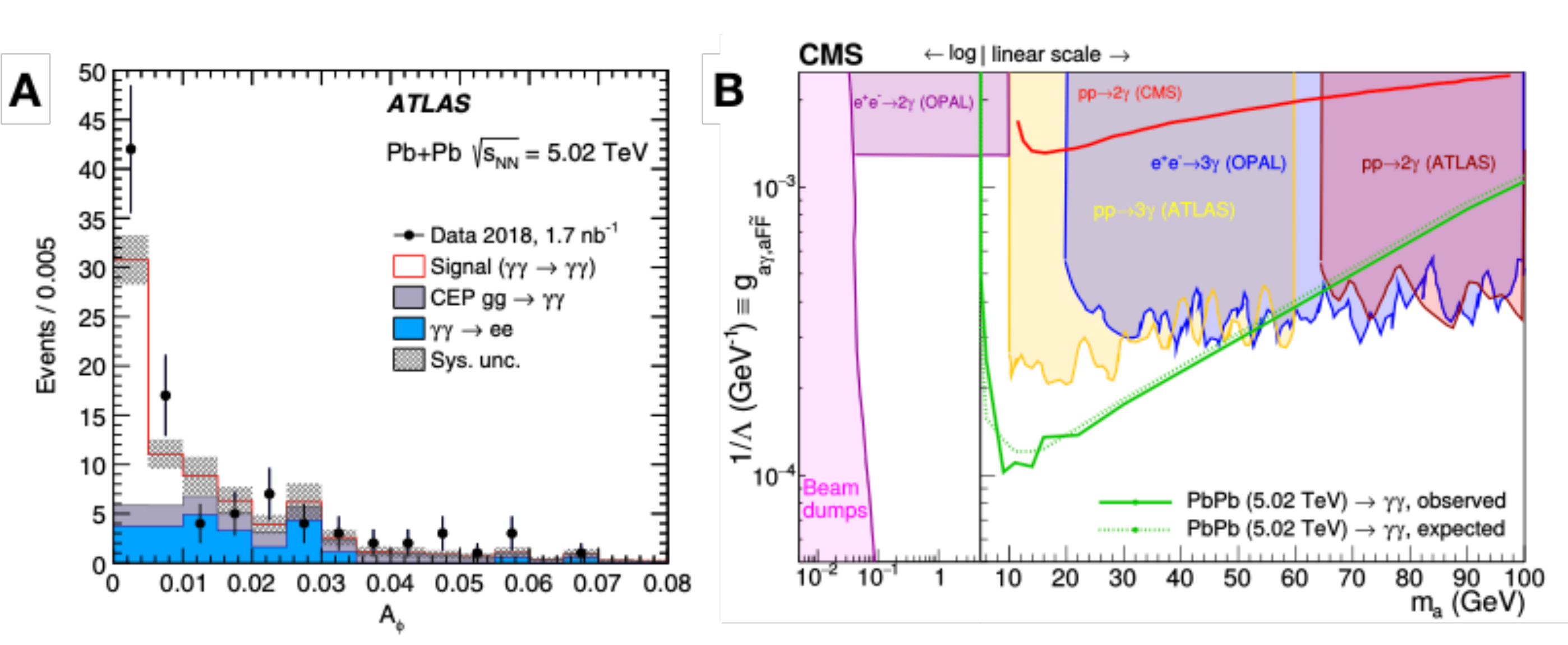}
\caption{(a.) Acoplanarity ($A_\phi$) distributions from ATLAS exclusive diphoton events~\cite{Aad:2019ock}, compared with expectations from the two primary backgrounds ($e^+ e^-$ and CEP) (b.) Upper limits set on axion-like particle production, assuming only coupling to photons~\cite{Sirunyan:2018fhl}.}
\label{fig:light-by-light-plots}
\end{figure}

The CMS result~\cite{Sirunyan:2018fhl} utilized similar techniques and addressed the same background contributions, but with a slightly larger fiducial space ($E_\mathrm{T} > 2$ GeV per photon, and $M_{\gamgam}>5$ GeV).  The lower integrated luminosity provided only 14 events in the signal region, with approximately $4\pm1$ background events expected, giving a significance of 3.7 standard deviations.
CMS also provided upper limits on axion-like particle production cross section, $\gamgam \rightarrow a \rightarrow \gamgam$, over a range $5<m_a<90$ GeV and translated these to limits on the photon-axion coupling $g_{a\gamma}$.  These are performed using calculations~\cite{Knapen:2016moh} based on two assumptions, the first being photon-only coupling  (shown in Fig.~\ref{fig:light-by-light-plots} (b)), where new limits are provided over $5<m_a<50$ GeV, and the other including hypercharge coupling (not shown) where new limits are set only in a more limited range $5<m_a<10$ GeV, just beyond the region explored by an earlier ATLAS $3\gamma$ analysis~\cite{Aad:2015bua}.

\subsection{Hadron production: $\gamongam \rightarrow X$}

Many early papers discussed two-photon production of mesons and meson pairs. Unfortunately, however, the cross-sections and production rates are rather small \cite{Bertulani:2005ru}, and even the benchmark $\gamgam \rightarrow f_2(1270)$ process has not yet been observed. In many cases, the calculated production rates due to feed-down from photoproduction are larger than the two-photon production rates.  For example, the rate for coherent photoproduction of the $J/\psi$ followed by $J/\psi\rightarrow \eta_c\gamma$ is considerably larger than for $\gamma\gamma\rightarrow \eta_c$ \cite{Klein:2018ypk}.  The photon is too soft to be seen in collider detectors, and the $\eta_c$ kinematics from the two processes are very similar. 

\section{Future prospects}

The current heavy-ion datasets (A+A and p+A) have not been fully utilized to explore all possible aspects of the physics discussed in here.  One can expect further improvements for processes involving higher photon energies based on the full integrated luminosity in the LHC Run 2 dataset (up to about 2 nb$^{-1}$).  However, LHC Runs 3 and 4, while keeping the same beam energies (or perhaps increasing by 7.6\%) will collect nearly an order of magnitude more integrated luminosity for both ion-ion and proton-ion collisions~\cite{Citron:2018lsq}.  A run with intermediate mass ions is also possible; this allows higher UPC luminosity due to the substantially-reduced BFPP and Coulomb excitation cross sections. 

CERN's planned Future Circular Collider or the proposed Chinese SPPC \cite{CEPC-SPPCStudyGroup:2015csa} would allow for a greatly expanded scope of measurements. They will have a higher beam energy by nearly an order of magnitude, which should allow for studies of photoproduction of top, along with photoproduction and two-photon physics studies of vector bosons, and improved opportunities to search for beyond-standard-model physics. 

The next generation of RHIC experiments, with an updated STAR detector and sPHENIX, a newly-built collider experiment focusing on extremely high data rates, will have access to enhanced statistics on exclusive dilepton, vector meson, and possibly even UPC jet physics.  With the new generation of high-precision vertex detectors, studies of photoproduction of open charm should also be possible.

UPC photoproduction studies will be complemented by photoproduction ($Q^2\approx 0$) and electroproduction ($Q^2>0$) studied at a future electron-ion collider (EIC), such as the planned U. S. electron-ion collider at Brookhaven National Laboratory \cite{Accardi:2012qut} and CERN's LHeC \cite{AbelleiraFernandez:2012cc}.  Although, per Table 1, most of the EIC designs reach lower $W_{\gamma p}$ than the LHC, they offer considerably higher luminosity.  More importantly, electrons can radiate photons over a wide range of $Q^2$, and this $Q^2$ can be determined from observing the scattered electron, no matter what reaction occurred.  The wide $Q^2$ range corresponds to a broad dipole  size distribution of dipoles, no matter what the reaction. This will allow us to better probe the gluon distribution in nuclei on different length scales.  An EIC will also allow us to study two-photon physics where one of the photons is highly virtual. 

\begin{summary}[SUMMARY POINTS]
\begin{enumerate}
\item Ultra-peripheral collisions are the energy frontier for photonuclear and two-photon physics, with the LHC reaching higher center of mass energies than any existing alternative. 
\item High-precision $J/\psi$ photoproduction data is consistent with moderate shadowing scenarios. 
\item Two-photon processes involving dilepton and two-photon final states are providing new insights from QED and potential access to physics beyond the standard model
\end{enumerate}
\end{summary}

\begin{issues}[FUTURE ISSUES]
\begin{enumerate}
\item Existing and future RHIC and LHC data can be probed to study a wide range of photoproduction and two-photon reactions, particularly including studies of a wider range of final state mesons, all species of dileptons and heavy quarks
\item Precision studies of benchmark photoproduction of multiple heavy mesons, with and without forward neutron topological selections, will be needed to improve our knowledge of nuclear structure, to test different saturation and colored-glass condensate models and measure nuclear shadowing. 
\item A future high-energy FCC-hh collider will extend photoproduction and two-photon physics studies by a factor of 7 in energy, opening the door to study top quark physics, probe parton distributions down to $x \approx 10^{-7}$, observe $\gamongam \rightarrow H$ and provide complementary probes to physics beyond the standard model.
\item A future electron-ion collider will be able to make high-precision complementary studies of many photoproduction channels over a wide range of $Q^2$.
\end{enumerate}
\end{issues}

\section*{DISCLOSURE STATEMENT}
The authors are not aware of any affiliations, memberships, funding, or financial holdings that
might be perceived as affecting the objectivity of this review. 

\section*{ACKNOWLEDGMENTS}

The authors thank their STAR, ALICE and ATLAS colleagues. This work was supported by
the U.S. Department of Energy, Office of Science, Office of Nuclear Physics, under contract number DE-AC02-05CH11231 and DE-SC0012704.  

\bibliographystyle{ar-style5}
\bibliography{refs} 

\end{document}